\documentclass[12p]{article}
 \usepackage{dcolumn}% Align table columns on decimal point
   \usepackage{threeparttable}
    \newcommand{\thickhline}{\noalign{\hrule height 0.8pt}}
\usepackage{latexsym}
\usepackage{cite}
\usepackage{color}
\usepackage{amsmath}
\usepackage{epsfig}
\usepackage{hyperref}
 \usepackage{subfigure}
\newcommand{\ortala}[1]{\begin{center}#1\end{center}}

\newcommand{\sandd}[1]{\left\langle #1\right\rangle}
\newcommand{\sanddr}[1]{\left\langle\left\langle #1\right\rangle\right\rangle_r}

\newcommand{\integ}[3]{{{\underset{#1 }{\overset{#2}{\displaystyle\int}}}#3}}

\newcommand{\summ}[3]{{{\underset{#1 }{\overset{#2}{\displaystyle\sum}}}#3}}
\newcommand{\prodd}[3]{{{\underset{#1
}{\overset{#2}{\displaystyle\prod}}}#3}}

\newcommand{\re}[1]{(\ref{#1})}

\newcommand{\eq}[2]{\begin{equation}\label{#1}  #2\end{equation}}

\newcommand{\paran}[1]{\left(#1\right)}

\newcommand{\sch}[1]{Schrodinger}

\linethickness{0.6pt}

\begin{document}
%\tableofcontents

\ortala{\textbf{An improved effective field theory formulation of spin-1
Ising systems with arbitrary coordination number $z$}}

\ortala{\"Umit Ak\i nc\i\footnote{umit.akinci@deu.edu.tr}}
\ortala{\textit{Department of Physics, Dokuz Eyl\"ul University, TR-35160 Izmir, Turkey}}

\section{Abstract}\label{abstract}

An improved  unified formulation based on the effective field theory is introduced for a
spin-1 Ising model with nearest neighbor interactions with arbitrary coordination
number $z$. Present formulation is capable of calculating all the multi-spin correlations systematically in a representative manner,
as well as its single site counterparts in the system and gives much better results
for critical temperature, in comparison with the other works in the literature.  The formulation
can be easily used to various kinds of spin-1 Ising models,
as long as the system contains only the nearest neighbor interactions as spin-spin interactions. Keywords:
\textbf{spin-1 Ising model ; Effective field theory ;Correlation functions; Bond diluted Ising system; Random field distributions, Crystal field dilution
}

\section{Introduction}\label{introduction}

Spin-1 Blume-Capel (BC) model \cite{ref1,ref2}  is one of the most extensively studied models in statistical mechanics. Beside
the various real magnetic systems, the model or some extensions can describe many physical systems such as ternary
alloys and multicomponent fluids such as $^3He-^4He$ mixtures\cite{ref3}. Although the BC model is a very simple model, it exhibits many
multicritical phenomena such as first (second) order order-disorder phase transitions, tricritcal points. On
the other hand by inclusion of random distributions of crystal field and/or external longitudinal magnetic field or
bond dilution of the BC model, makes the model more valuable. For example, BC model with random crystal field can describe $^3He-^4He$ mixtures in
a random media i.e. aerogel\cite{ref4}. Taking into account these randomness effects will change phase diagrams of the model drastically, then model would exhibit
richer multicritical phenomena.

The spin-1 BC model was studied in different lattices by a variety of methods such as two spin cluster approximation \cite{ref6},
Bethe lattice approximation \cite{ref7}, series expansion method \cite{ref8,ref9}, cluster variation method \cite{ref10,ref11},
Monte Carlo simulations \cite{ref12,ref13,ref14,ref15}, renormalization group \cite{ref16,ref17,ref18} and effective field theory (EFT) \cite{ref19,ref20,ref21}.
In order to simplify the application and improve the results of EFT, Kaneyoshi introduced differential operator technique
\cite{ref22,ref23,ref24,ref25,ref26} and Du introduced the expanded Bethe-Peierls approximation (BPA) \cite{ref27,ref28,ref29}.

On the other hand there exist several works dealing with the BC model with random crystal field and bond dilution.
Among these works, \cite{ref30,ref31,ref32,ref33,ref34} examined the random crystal field system with EFT.
Some other applied techniques  about this system are, mean field theory  \cite{ref35,ref36,ref37,ref38,ref39,ref40, ref41}, cluster variation method\cite{ref42},
 Bethe lattice approximation\cite{ref43}, finite cluster approximation\cite{ref44,ref45}, Monte Carlo simulation\cite{ref46},  pair approximation\cite{ref47} and
 renormalization group method\cite{ref48}. Also the bond dilution problem in (transverse field) BC model was investigated with similar methods such as
 EFT \cite{ref49,ref50,ref51,ref52,ref53,ref54,ref56,ref57,ref58,ref59,ref60,ref62,ref63}, finite cluster approximation\cite{ref64,ref65,ref66,ref67,ref68}, Monte Carlo simulations\cite{ref69},
 series expansion\cite{ref71}, two spin cluster approximation\cite{ref72,ref73,ref74} and renormalization group method\cite{ref75}.

As seen in this very short literature about the spin-1 BC model and some of its extensions, the EFT method for solutions
of this model has an important place in the literature. Thus, the aim of this work is to introduce
an improved  general effective field formulation for the spin-1 BC model and its variants, which is capable of calculating all multi site correlations of the
system, in contrast to the other EFT formalisms in the related literature.

A typical EFT method or finite cluster approximation start by choosing a finite representative cluster and using exact spin identity (e.g. given in\cite{ref76})
 which gives the thermal average of magnetization. In the process of evaluating this identity in the cluster, multi site spin correlations appear. Almost all of the works
in the literature treat these multi site spin correlations by using a decoupling approximation (DA) which completely neglects
them. Recently we have treated these multi site spin correlations in the spin-1 BC model \cite{ref77,ref78,ref79} for the specific lattices and showed
that taking into account these correlations considerably improves the results. Beside the improvement of the results, calculation of these correlations
allows us to obtain some thermodynamic quantities more accurately (e.g internal energy, specific heat).

The aim of this work is to obtain a general EFT formulation via treating the multi site spin correlations which is valid for spin-1 BC model and its variants for
any arbitrary coordination number, as long as the system contains only nearest neighbor interactions as spin-spin interactions. For this purpose, the paper is organized
as follows: In Sec. \ref{model} we present this general  formulation. As an application of our formulation we give the results of
the pure, bond diluted and crystal field diluted spin-1 BC model on  three dimensional lattices in Sec. \ref{results}, and finally Sec. \ref{conclusion} contains our conclusions.

\section{Model and Formulation}\label{model}
The Hamiltonian of our system is given by
\eq{denk1}{\mathcal{H}=-\summ{<i,j>}{}{J_{ij}s_is_j}-\summ{i}{}{D_is_i^2}-\summ{i}{}{H_is_i}}
where $s_i$ is the $z$ component of the spin and it takes the values
$s_i=0,\pm 1$ for the spin-1 system, $J_{ij}=J_{ji}>0$ is the ferromagnetic exchange
interaction between spins $i$ and $j$, $D_i$ is the crystal field,
$H_i$ is the external longitudinal magnetic field at a lattice site $i$. The first
summation in Eq. \re{denk1} is over the nearest-neighbor pairs of
spins and the other summations are over the all lattice sites.
$J_{ij}$, $D_i$ and $H_i$ may be given with certain distributions or
they can have the same values for all pairs/sites, i.e $J_{ij}=J, D_i=D,
H_i=H$.

We consider a lattice which has N identical spins arranged. We define a cluster on the lattice which
consists of a central spin labeled $s_0$, and $z$ perimeter spins being the nearest neighbors of the central spin. The nearest-neighbor spins are in an effective field produced by the outer spins, which can be determined by the condition that the thermal average of the central spin is equal to that of its nearest-neighbor spins.

We start with spin identities for spin-1 Ising system \cite{ref76} which are
given by
\eq{denk2}{\sanddr{\{f_i\}s_i}=\sanddr{\{f_i\}\frac{Tr_i s_i
\exp{\paran{-\beta \mathcal{H}_i}}}{Tr_i\exp{\paran{-\beta
\mathcal{H}_i}}}}}
\eq{denk3}{\sanddr{\{f_i\}s_i^2}=\sanddr{\{f_i\}\frac{Tr_i s_i^2
\exp{\paran{-\beta \mathcal{H}_i}}}{Tr_i\exp{\paran{-\beta
\mathcal{H}_i}}}}} where $Tr_i$ is the partial trace over the site
$i$, $\beta=1/\paran{k_B T}$, $k_B$ is Boltzmann constant and $T$ is
the temperature, $\{f_i\}$ is any function as long as it is not a
function of the site $i$, $\mathcal{H}_i$ is the part of the
Hamiltonian which includes all contributions associated
with the site $i$. While the inner average brackets stands for the thermal average, outer average brackets (with subscript $r$) stands
for the random configurational average which is necessary for including random distribution effects.

Let us denote $s_i^2$ as $q_i$. Performing partial traces in Eqs. \re{denk2} and \re{denk3} with $\mathcal{H}_0=-s_0\summ{\delta=1}{z}{J_{0\delta}s_\delta}$ $-q_0 D_0-s_0H_0$ for central spin ($s_0$) and with
$\mathcal{H}_\delta=-s_\delta J_{0\delta}s_0 - s_\delta \gamma-q_\delta D_\delta-s_\delta H_\delta $ for perimeter spin ($s_\delta$) then with using differential operator technique
\cite{ref24}  will yield
\eq{denk4}{\sanddr{\{f_0\}s_0}=\sanddr{\{f_0\}\prodd{\delta=1}{z}{\left[1+s_\delta \sinh{\paran{J_{0\delta}\nabla}}+q_\delta\paran{\cosh{\paran{J_{0\delta}\nabla}}-1}\right]}}F_1(x)|_{x=0}}

\eq{denk5}{
\sanddr{\{f_0\}q_0}=\sanddr{\{f_0\}\prodd{\delta=1}{z}{\left[1+s_\delta \sinh{\paran{J_{0\delta}\nabla}}+q_\delta\paran{\cosh{\paran{J_{0\delta}\nabla}}-1}\right]}}F_2(x)|_{x=0}
}
\eq{denk6}{\sanddr{\{f_\delta\}s_\delta}=\sanddr{\{f_\delta\}\left[1+s_0 \sinh{\paran{J_{0\delta}\nabla}}+q_0\paran{\cosh{\paran{J_{0\delta}\nabla}}-1}\right]}F_1(x+\gamma)|_{x=0}}
\eq{denk7}{\sanddr{\{f_\delta\}q_\delta}=\sanddr{\{f_\delta\}\left[1+s_0 \sinh{\paran{J_{0\delta}\nabla}}+q_0\paran{\cosh{\paran{J_{0\delta}\nabla}}-1}\right]}F_2(x+\gamma)|_{x=0}}
where $\nabla=\partial/\partial x$ differential operator, $\gamma=(z-1)h$ and $h$ is the effective field per spin. The effect of the differential operator on a function is given by
\eq{denk8}{\exp{\paran{a\nabla}}F_j\paran{x}=F_j\paran{x+a}, \quad j=1,2} with any constant  $a$.

Eqs. \re{denk4}-\re{denk7} are our fundamental equations and the functions in the equations are given by via,
\eq{denk9}{
G_1(x,D_i,H_i)=\frac{2\sinh{\left[\beta\paran{x+H_i}\right]}}{2\cosh{\left[\beta\paran{x+H_i}\right]}+\exp{\paran{-\beta D_i}}}, \quad
G_2(x,D_i,H_i)=\frac{2\cosh{\left[\beta\paran{x+H_i}\right]}}{2\cosh{\left[\beta\paran{x+H_i}\right]}+\exp{\paran{-\beta D_i}}}
}
as
\eq{denk10}{
F_j(x)=\integ{}{}{dD_idH_i P_D\paran{D_i}P_H\paran{H_i}G_j\paran{x,D_i,H_i}, \quad j=1,2}
} where $P_D\paran{D_i}$ and $P_H\paran{H_i}$ are probability distributions of the crystal field and longitudinal magnetic field respectively.

We can obtain the central spin magnetization and the quadrupolar moment from
Eqs. \re{denk4} and \re{denk5} by letting $\{f_0\}=1$ and expanding right hand sides of them.
Both expansions have multi site spin correlations like
$\sanddr{s_1s_2\ldots q_{k-1}q_{k}}$ where $k\le z$. The main problem
here is to handle these correlations. These correlations are treated
with DA in most of the works in
literature due to the mathematical difficulties. In DA these
correlations are treated according to
\eq{denk11}{\sanddr{s_1s_2\ldots q_{k-1}q_{k}}=\sanddr{s_1}
\sanddr{s_2} \ldots \sanddr{q_{k-1}} \sanddr{q_{k}}} i.e neglecting all
the multi site spin correlations. But these correlations can be obtained
by choosing suitable form of $\{f_\delta\}$ in Eqs. \re{denk6} or \re{denk7}.
This is the main point of this work. If we choose a proper form of $\{f_\delta\}$
in Eqs. \re{denk6} or \re{denk7}, this time other multi site correlations
appear on the right of sides of Eqs. \re{denk6} and \re{denk7}, which include the central
site. By the same way, these can be obtained from Eqs. \re{denk4} or
\re{denk5} with a suitable $\{f_0\}$. At the end of this
process we can obtain a set of linear equations which includes
these correlations as unknowns. For a given set of Hamiltonian parameters,
we can obtain the effective field (which represents the effect of the
outer $(z-1)$ spins) by usual condition that
\eq{denk12}{\sanddr{s_{0}}=\sanddr{s_{\delta}}.} Once effective field
$h$ is determined for a given parameters, we can solve the system of linear
equations and obtain the whole correlations of the system for
given parameters. Since the effective field is very small in the
vicinity of the phase transition point, the critical temperature can be
obtained by letting $\gamma\rightarrow 0$.

In the formulation presented here, instead of deriving all correlations in the system, we construct representative correlations.
Let us denote the  correlation $\sanddr{s_1s_2\ldots s_p q_{p+1}q_{p+2}\ldots q_{n}}$ as
\eq{denk13}{\sanddr{s_\delta^{(p)}q_\delta^{(n-p)}}=\sanddr{s_1s_2\ldots s_p q_{p+1}q_{p+2}\ldots q_{n}}.}

For instance, if we choose
$n=2,p=1$ in Eq. \re{denk13} for the system with $z=3$, the correlation $\sanddr{s_\delta^{(1)}q_\delta^{(1)}}=\sanddr{s_1q_2}$ represents all the terms
$\sanddr{s_1q_2},\sanddr{s_1q_3},\sanddr{s_2q_1},\sanddr{s_2q_3},$
$\sanddr{s_3q_1},\sanddr{s_3q_2}$. This means that our solutions are under the assumption of that all these correlations are
equal to each other. In general, we make an assumption that
all $n$ perimeter site correlations which includes $p$ number of $s_\delta$ and $n-p$ number of $q_\delta$ are equal to each other and we represent them by
$\sanddr{s_\delta^{(p)}q_\delta^{(n-p)}}$ as given in Eq. \re{denk13}. The same assumption holds for the correlations which include central site and $n$ perimeter sites,
i.e all correlations which include $s_0$ and $p$ number of $s_\delta$ and $n-p$ number of $q_\delta$ are equal to each other and all correlations which include $q_0$ and $p$ number of
$s_\delta$ and $n-p$ number of $q_\delta$ are equal to each other, and we represent
them as $\sanddr{s_0s_\delta^{(p)}q_\delta^{(n-p)}}$ and $\sanddr{q_0s_\delta^{(p)}q_\delta^{(n-p)}}$, respectively.

Deriving correlations simply based on choosing $\{f_0\}$ in Eqs. \re{denk4} or \re{denk5}, or  $\{f_\delta\}$ in Eqs. \re{denk6} or \re{denk7} and chosen $\{f_0\}$ and $\{f_\delta\}$, will change the operator in the average brackets on the right hand sides of them.
For example, by the choice of $\{f_0\}=s_k$ in Eq. \re{denk4} one applies $s_k$ to the expression $\sanddr{s_0}$. Applying
$s_1$ to the expression $\sanddr{s_0}$ will create the correlation $\sanddr{s_0s_1}$. After then applying $s_2$ to the $\sanddr{s_0s_1}$ will create $\sanddr{s_0s_1s_2}$ correlation,
and so on. Thus, we get the correlation $\sanddr{s_0s_1\ldots s_z}$  with successive applications of $s_k$  to the
correlation $\sanddr{s_0s_1\ldots s_{k-1}}$ ($k=1,2,\ldots,z$) in $z$ steps.

%Derivation process in given sequential orders, based on applying some $s_k$ to some correlation expression, will cause modifying the operator in the right hand side
%of the correlation expression.

Since our starting points of the derivation process are Eqs. \re{denk4}-\re{denk7}  we can write any correlation in terms of the related
operator which is applied to the related function, as in expressions Eqs. \re{denk4}-\re{denk7}.  Since the operators on the right hand sides of Eqs. \re{denk4} and \re{denk5} are
the same, we will use the same operator for getting the correlations e.g. $\sanddr{s_0s_1s_2}$ and $\sanddr{q_0s_1s_2}$. The same reasoning also holds for
the derivation of the correlations from Eqs. \re{denk6} and \re{denk7}. Thus, instead of deriving all correlations separately in sequential orders, we will concentrate on the evolution of
these operators in that sequential orders. Let us denote the operator on the right hand sides of  Eqs. \re{denk4} and \re{denk5} for $\{f_0\}=1$ as $\Theta_{0,0}$  and
the operator on the right hand sides of  Eqs. \re{denk6} and \re{denk7} with $\delta=1,\{f_1\}=1$ as $\Phi_{0,0}$. With these operators we can write our fundamental
equations given in Eqs. \re{denk4}-\re{denk7} as follows:

\eq{denk14}{\sanddr{s_0}=\sanddr{\Theta_{00}}F_1(x)|_{x=0}}
\eq{denk15}{\sanddr{s_1}=\sanddr{\Phi_{00}}F_1(x+\gamma)|_{x=0}}

\eq{denk16}{\sanddr{q_0}=\sanddr{\Theta_{00}}F_2(x)|_{x=0}}
\eq{denk17}{\sanddr{q_1}=\sanddr{\Phi_{00}}F_2(x+\gamma)|_{x=0}}
where the operators are given as
\eq{denk18}{\Theta_{00}=\prodd{\delta=1}{z}{\left[1+s_\delta \sinh{\paran{J_{0\delta}\nabla}}+q_\delta\paran{\cosh{\paran{J_{0\delta}\nabla}}-1}\right]}}
\eq{denk19}{\Phi_{00}=\left[1+s_0 \sinh{\paran{J_{0\delta}\nabla}}+q_0\paran{\cosh{\paran{J_{0\delta}\nabla}}-1}\right].}

Let us derive correlations in sequential orders given below. At each step of the derivation, the operators evolve in the sequences as given below:

\eq{denk20}{
\begin{array}{ccccccc}
\sanddr{s_1}&\rightarrow& \sanddr{s_1s_2}&\rightarrow& \ldots &\rightarrow& \sanddr{s_1s_2\ldots s_z}\\
\Phi_{00}&\rightarrow& \Phi_{10}&\rightarrow& \ldots &\rightarrow& \Phi_{{z-1,0}}
\end{array}
}
\eq{denk21}{
\begin{array}{ccccccccc}
\sanddr{s_0}&\rightarrow&\sanddr{s_0s_1}&\rightarrow& \sanddr{s_0s_1s_2}&\rightarrow& \ldots &\rightarrow& \sanddr{s_0s_1s_2\ldots s_z}\\
\Theta_{00}&\rightarrow&\Theta_{10}&\rightarrow& \Theta_{20}&\rightarrow& \ldots &\rightarrow& \Theta_{z0}
\end{array}
}
\eq{denk22}{
\begin{array}{ccccccc}
\sanddr{s_1s_2\ldots s_{k+1}}&\rightarrow& \sanddr{s_1s_2\ldots s_{k}q_{k+1}}&\rightarrow& \ldots &\rightarrow& \sanddr{s_1s_2\ldots s_{k-m+1}q_{k-m+2}\ldots q_{k+1}}\\
\Phi_{k0}&\rightarrow& \Phi_{k1}&\rightarrow& \ldots &\rightarrow& \Phi_{{km}}
\end{array}
}
\eq{denk23}{
\begin{array}{ccccccc}
\sanddr{s_0s_1s_2\ldots s_k}&\rightarrow& \sanddr{s_0s_1s_2\ldots s_{k-1}q_k}&\rightarrow& \ldots &\rightarrow& \sanddr{s_0s_1s_2\ldots s_{k-m}q_{k-m+1}\ldots q_k}.\\
\Theta_{k0}&\rightarrow& \Theta_{k1}&\rightarrow& \ldots &\rightarrow& \Theta_{{km}}.
\end{array}
}

In Eqs. \re{denk22} and \re{denk23} we use the self spin identities of spin-1 Ising system during the derivation process, which are given as follows.
\eq{denk24}{s_iq_i=s_i,s_is_i=q_i,q_iq_i=q_i, \quad i=0,1,\ldots z.} According to Eq. \re{denk24}, applying $s_{l}$ to the correlation $\sanddr{s_1s_2\ldots s_{l}q_{l+1}\ldots q_k}$ will
produce the correlation $\sanddr{s_1s_2\ldots s_{l-1}q_{l}q_{l+1}\ldots q_k}$ where $l\le k$, which can be obtained by applying the $\sanddr{\Phi_{{k-1,k-l+1}}}$ operator to
the function $F_1\paran{x+\gamma}$ and taking the value of resultant expression at $x=0$.

Sequential orders given in Eqs. \re{denk20}-\re{denk23}, expose the recurrence relations for the operators as:
\eq{denk25}{
\begin{array}{lcl}
\Phi_{k,0}&=&s_{k+1}\Phi_{k-1,0}, \quad k=1,2,\ldots z-1\\
\Theta_{k,0}&=&s_k\Theta_{k-1,0}, \quad k=1,2,\ldots z\\
\Phi_{k,m}&=&s_{k-m+2}\Phi_{k,m-1}, \quad m=1,2,\ldots k\\
\Theta_{k,m}&=&s_{k-m+1}\Theta_{k,m-1}, \quad m=1,2,\ldots k.\\
\end{array}
}

In order to obtain these operators let us start with the operator $\Theta_{00}$ which is just the operators on the right hand sides of Eqs. \re{denk4} and \re{denk5} with $\{f_0\}=1$.
By using the assumption about the correlations which is explained above (under the Eq. \re{denk13}),
 we can write  this operator as
\eq{denk26}{
\Theta_{00}=\summ{n=0}{z}{}\summ{p=0}{n}{}A_{np}s_\delta^{(p)}q_\delta^{(n-p)}
}  where

\eq{denk27}{A_{np}=\paran{\begin{array}{c}z\\n\end{array}}\paran{\begin{array}{c}n\\p\end{array}}
\prodd{\delta=1}{p}{\sinh{\paran{J_{0\delta}\nabla}}}\prodd{\delta=p+1}{n}{\paran{\cosh{\paran{J_{0\delta}\nabla}}-1}}.}

However, obtaining $\Theta_{k0}$ from $\Theta_{k-1,0}$ by using Eq. \re{denk25} requires the determination of any term $s_ks_\delta^{(l)}q_\delta^{(m)}$ which
will appear on the right hand side of $\Theta_{k0}$. This requirement is also valid for any process for obtaining the operators by using the recurrence relations given
in Eq. \re{denk25}. At this stage we use
self spin correlations for the spin-1 Ising system given in Eq. \re{denk24}.
Hence  there exist three possibilities for the term $s_ks_\delta^{(l)}q_\delta^{(m)}$ according to Eq. \re{denk24}:
\eq{denk28}{
s_ks_\delta^{(l)}q_\delta^{(m)}=\left\{\begin{array}{lcl}s_\delta^{(l+1)}q_\delta^{(m)}&,& \quad k>l+m\\
s_\delta^{(l+1)}q_\delta^{(m-1)}&,& l<k\le l+m\quad \\
 s_\delta^{(l-1)}q_\delta^{(m+1)}&,& \quad 0<k \le l\\
\end{array}\right.
}

Now, one strategy is to obtain $\Theta_{k0}$ from Eq. \re{denk25} starting by Eq. \re{denk26} is making first few iterations and trying to capture the general expression for the
$\Theta_{k0}$. However we follow a slightly different strategy here. Each of the iterations in Eq. \re{denk25} takes the coefficient $A_{np}$ which is the multiplier
of some $s_\delta^{(l)}q_\delta^{(m)}$ and place it to another term $s_\delta^{(l^\prime)}q_\delta^{(m^\prime)}$ as a multiplier. We envision that the
iterations in Eq. \re{denk25}  will generate certain movement of the coefficients $A_{np}$ between the terms $s_\delta^{(l)}q_\delta^{(m)}$ as a multiplier of them. Then if we can track every coefficient
during the each step of the iterations, we can determine the place of any coefficient for that iteration step. Thus we can write an expression for any operator in any iteration step.

For instance let us take the
coefficient $A_{00}$ and track this coefficient during the
derivation in a sequential order given in Eq. \re{denk21}.
\eq{denk29}{A_{00}s_\delta^{(0)}q_\delta^{(0)}\rightarrow
A_{00}s_\delta^{(1)}q_\delta^{(0)}\rightarrow \ldots \rightarrow
A_{00}s_\delta^{(k)}q_\delta^{(0)}\rightarrow\ldots \rightarrow
A_{00}s_\delta^{(z)}q_\delta^{(0)}} Eq. \re{denk29} implies that, the
coefficient $A_{00}$ is placed as a multiplier of the term
$s_\delta^{(k)}q_\delta^{(0)}$ on the $k.th$ step of the derivation, i.e the operator expression $\Theta_{k0}$
has the term
$A_{00}s_\delta^{(k)}q_\delta^{(0)}=A_{00}s_1s_2\ldots s_k$ on the
right hand side. If we know the locations of the all $A_{np}$ coefficients at the $k_{th}$ step of the derivation we can get an expression for
$\Theta_{k0}$. This is the general strategy which is
used to obtain all the operators in Eq. \re{denk21}.

By inspection and taking into account Eq. \re{denk28}, in order to get an expression for the operator $\Theta_{k0}$, we can give the locations of the coefficients of the
$k_{th}$ step of the derivation in an order given in Eq. \re{denk21} as follows.
\eq{denk30}{
A_{00}\rightarrow s_\delta^{(k)}q_\delta^{(0)}.
}
For $n>0$
\eq{denk31}{
 A_{n0}\rightarrow \left\{\begin{array}{lcl}s_\delta^{(k)}q_\delta^{(n-k)}&,& \quad k\le n\\
s_\delta^{(k)}q_\delta^{(0)}&,& \quad k>n \\
\end{array}\right.
}
For $p\ne 0$ and even
\eq{denk32}{
A_{np}\rightarrow \left\{\begin{array}{lcl}s_\delta^{(p-k)}q_\delta^{(n-p+k)}&,& \quad k\le p/2\\
s_\delta^{(k)}q_\delta^{(n-k)}&,& p/2<k\le n\quad \\
s_\delta^{(k)}q_\delta^{(0)}&,& \quad k > n\\
\end{array}\right.
}
and for odd $p$
\eq{denk33}{
A_{np}\rightarrow \left\{\begin{array}{lcl}s_\delta^{(p-k)}q_\delta^{(n-p+k)}&,& \quad k\le (p+1)/2\\
s_\delta^{(k-1)}q_\delta^{(n-k+1)}&,& (p+1)/2<k\le n\quad \\
s_\delta^{(k-1)}q_\delta^{(1)}&,& \quad k > n\\
\end{array}\right.
}
Detailed derivation of Eqs. \re{denk30}-\re{denk33} is given in Section \ref{App_a}. From Eqs. \re{denk30}-\re{denk33} we can get an expression for the operator $\Theta_{k0}$ as
$$
\Theta_{k,0}=\paran{D_k^{(1)}+D_k^{(2)}}s_\delta^{(k)}q_\delta^{(0)}+D_k^{(3)}s_\delta^{(k-1)}q_\delta^{(1)}+
\summ{n=k}{z}{\paran{A_{n0}+D_{nk}^{(4)}}s_\delta^{(k)}q_\delta^{(n-k)}}
$$
\eq{denk34}{
+\summ{n=k}{z}{D_{nk}^{(5)}s_\delta^{(k-1)}q_\delta^{(n-k+1)}}
+\summ{n=k,}{z}{}\summ{p=2k-1}{n}{}A_{np}s_\delta^{(p-k)}q_\delta^{(n-p+k)}
}
where $k=1,2,\ldots,z$ and the coefficients are given by

$$
D_k^{(1)}=\summ{n=0}{k-1}{A_{n0}},\quad D_k^{(2)}=\summ{n=1}{k-1}{}\summ{p=2}{n^{\prime\prime}}{}A_{np},
\quad D_k^{(3)}=\summ{n=1}{k-1}{}\summ{p=1}{n^{\prime}}{}A_{np}
$$
\eq{denk35}{
D_{nk}^{(4)}=\summ{p=2}{2k-1^{\prime\prime}}{}A_{np},\quad D_{nk}^{(5)}=\summ{p=1}{2k-2^{\prime}}{}A_{np}.
}The $\prime$ sign on the upper limits of the sums in \re{denk35} indicate that, sum runs over the odd indexes while
the $\prime\prime$ sign on the upper limits of the sums indicate that sum runs over the even indexes.

In the sequential order given in  Eq. \re{denk23}, we can obtain the operator $\Theta_{km}$ after a few iterations by using Eq. \re{denk28} , and it is given by
$$
\Theta_{k,m}=\paran{D_{k}^{(1)}+D_{k}^{(2)}}s_\delta^{(k-m)}q_\delta^{(m)}+
D_{k}^{(3)}s_\delta^{(k-m+1)}q_\delta^{(m-1)}
$$
$$
+\summ{n=k}{z}{\paran{A_{n0}+D_{nk}^{(4)}}s_\delta^{(k-m)}q_\delta^{(n-k+m)}}
$$
\eq{denk36}{
+\summ{n=k}{z}{\paran{A_{n,2k-1}+D_{nk}^{(5)}} s_\delta^{(k-m+1)}q_\delta^{(n-k+m-1)}}
+\summ{n=k,}{z}{}\summ{p=2k}{n}{}A_{np} s_\delta^{(p-k-m)}q_\delta^{(n-p+k+m)}
}
where $k=1,2,\ldots,z$; $m=1,2,\ldots k$.

%% Detailed derivation of  Eqs. \re{denk34} and \re{denk36} is given in Appendix B.

From  Eqs. \re{denk6} and \re{denk7} we can see the form of the operator $\Phi_{00}$ defined in Eqs. \re{denk15} and \re{denk17} as:
\eq{denk37}{\Phi_{00}=B_0+B_{1}s_0+B_{2}q_0}
where
\eq{denk38}{
\begin{array}{lcl}
B_{0}&=&1\\
B_{1}&=&\sinh{\paran{J_{0\delta}\nabla}}\\
B_{2}&=&\cosh{\paran{J_{0\delta}\nabla}}-1.
\end{array}
}
If we use the first line of the Eq. \re{denk25} by taking into account Eq. \re{denk28}, we can obtain the operator $\Phi_{k0}$ after a few iterations as,
\eq{denk39}{\Phi_{k0}=B_0s_\delta^{(k)}q_\delta^{(0)}+B_{1}s_0s_\delta^{(k)}q_\delta^{(0)}+
B_{2}q_0s_\delta^{(k)}q_\delta^{(0)}}
where $k=0,1,\ldots,z-1$. After then, by the same way, using the third line of the Eq. \re{denk25} we get
\eq{denk40}{
\Phi_{km}=B_0s_\delta^{(k-m)}q_\delta^{(m)}+B_{1}s_0s_\delta^{(k-m)}q_\delta^{(m)}+
B_{2}q_0s_\delta^{(k-m)}q_\delta^{(m)}
}
where $k=0,1,\ldots,z-1$ and $m=0,1,\ldots,k$. By using Eqs. \re{denk24} and \re{denk40} we can obtain the operator $s_0\Phi_{km}$ and $q_0\Phi_{km}$ as
\eq{denk41}{
\begin{array}{lcl}
s_0\Phi_{km}&=&(B_0+B_{2})s_0s_\delta^{(k-m)}q_\delta^{(m)}+B_{1}q_0s_\delta^{(k-m)}q_\delta^{(m)}\\
q_0\Phi_{km}&=&(B_0+B_{2})q_0s_\delta^{(k-m)}q_\delta^{(m)}+B_{1}s_0s_\delta^{(k-m)}q_\delta^{(m)}\\
\end{array}
} where $k=0,1,\ldots,z-1$, $m=0,1,\ldots,k$ and this completes the process of derivation of operators. Note that the definition given in Eq. \re{denk40} covers
the definition given in Eq. \re{denk39} with $m=0$ in it, but the same thing does not hold  Eq. \re{denk36} for Eq. \re{denk34} i.e the operator $\Theta_{k,0}$ have
to be calculated from Eq. \re{denk34}.

Now, since the operators and functions are well defined, we can construct the correlations. But there are more than one way to constructing any correlation. Thus we
have to determine an appropriate expressions from the equalities below, for deriving correlations.

The number of $z/2(z+3)-2$ correlations of the type $\sanddr{s_\delta^{(k-m)}q_\delta^{(m)}}$ ($k=2,3,\ldots,z$ and $m=0,1,\ldots,k$) can be obtained by starting from Eq. \re{denk15} or \re{denk17} as

\eq{denk42}{
\sanddr{s_\delta^{(k-m)}q_\delta^{(m)}} =
\sanddr{\Phi_{k-1,m}}F_1\paran{x+\gamma}|_{x=0}, m=0,1,\ldots,k-1
}

\eq{denk43}{
\sanddr{s_\delta^{(k-m)}q_\delta^{(m)}} =
\sanddr{\Phi_{k-1,m-1}}F_2\paran{x+\gamma}|_{x=0}, m=1,2,\ldots,k
} respectively, where $k=2,3,\ldots,z$. In order to obtain the complete set of correlations of this type, we can not use only one of Eqs. \re{denk42} and \re{denk43} due to
the restrictions on $m$ values of them.

Since the operator $\Theta_{k,m}$ defined in Eq. \re{denk36} can not have $m=0$ value,  the correlations of the type $\sanddr{s_0s_\delta^{(k)}}$ are obtained separately
from the correlations of the type $\sanddr{s_0s_\delta^{(k-m)}q_\delta^{(m)}}$ ($m\ne 0$). $z$ number of correlations of the type $\sanddr{s_0s_\delta^{(k)}}$ ($k=1,2,\ldots,z$) can
be obtained by starting from Eqs. \re{denk14} or \re{denk15} and they are given respectively as
\eq{denk44}{
\sanddr{s_0s_\delta^{(k)}} =
\sanddr{\Theta_{k,0}}F_1\paran{x}|_{x=0}
}
\eq{denk45}{
\sanddr{s_0s_\delta^{(k)}} =
\sanddr{s_0\Phi_{k-1,0}}F_1\paran{x+\gamma}|_{x=0}.
}

$z/2(z+1)$ number of  correlations of the type $\sanddr{s_0s_\delta^{(k-m)}q_\delta^{(m)}}$ ($k=1,2,\ldots,z$ and $m=1,2,\ldots,k$) can be obtained by
starting from  Eqs. \re{denk14} or \re{denk17} and they are given as

\eq{denk46}{\sanddr{s_0s_\delta^{(k-m)}q_\delta^{(m)}} =
\sanddr{\Theta_{k,m}}F_1\paran{x}|_{x=0}}

\eq{denk47}{\sanddr{s_0s_\delta^{(k-m)}q_\delta^{(m)}} =
\sanddr{s_0\Phi_{k-1,m-1}}F_2\paran{x+\gamma}|_{x=0}}
respectively.

In order to get the complete set of correlations of the type $\sanddr{s_0s_\delta^{(k-m)}q_\delta^{(m)}}$ including $m=0$ we can also
use
\eq{denk48}{\sanddr{s_0s_\delta^{(k-m)}q_\delta^{(m)}} =
\sanddr{s_0\Phi_{k-1,m}}F_1\paran{x+\gamma}|_{x=0}, m=0,1,\ldots,k-1}
with Eq. \re{denk47}. For $k=1,2,\ldots,z$, either Eq. \re{denk48} with $m=0$ and Eq. \re{denk47} with $m=1,2,\ldots k$ or
Eq. \re{denk47} with $m=k$ and Eq. \re{denk48} with $m=0,1,\ldots,k-1$ will generate number of $z/2(z+3)$ correlations which are the complete set of correlations including $s_0$.

Number of $z$ correlations of the type $\sanddr{q_0s_\delta^{(k)}}$ ($k=1,2,\ldots,z$) can be obtained by starting from Eqs. \re{denk15} or \re{denk16} and they are given respectively as
\eq{denk49}{\sanddr{q_0s_\delta^{(k)}} =
\sanddr{q_0\Phi_{k-1,0}}F_1\paran{x+\gamma}|_{x=0}}
\eq{denk50}{\sanddr{q_0s_\delta^{(k)}} =
\sanddr{\Theta_{k,0}}F_2\paran{x}|_{x=0}}

Number of $z/2(z+1)$ correlations of the type $\sanddr{q_0s_\delta^{(k-m)}q_\delta^{(m)}}$ ($k=1,2,\ldots,z$  and $m=1,2,\ldots,k$) can be obtained by starting from
Eqs. \re{denk16} or \re{denk17} and they are given as

\eq{denk51}{\sanddr{q_0s_\delta^{(k-m)}q_\delta^{(m)}} =
\sanddr{\Theta_{k,m}}F_2\paran{x}|_{x=0}}

\eq{denk52}{\sanddr{q_0s_\delta^{(k-m)}q_\delta^{(m)}} =
\sanddr{q_0\Phi_{k-1,m-1}}F_2\paran{x+\gamma}|_{x=0}}
respectively. Again, the complete set of correlations of the type $\sanddr{q_0s_\delta^{(k-m)}q_\delta^{(m)}}$ including $m=0$ can be obtained from
\eq{denk53}{\sanddr{q_0s_\delta^{(k-m)}q_\delta^{(m)}} =
\sanddr{q_0\Phi_{k-1,m}}F_1\paran{x+\gamma}|_{x=0}, m=0,1,\ldots,k-1}
with Eq. \re{denk52}. Due to the restrictions of $m$ in Eq. \re{denk52} and Eq. \re{denk53}, we can obtain the complete set of correlations which include
$q_0$, from  Eq. \re{denk53} with $m=0$ and \re{denk52} with $m=1,2,\ldots k$ or
\re{denk52} with $m=k$ and \re{denk53} with $m=0,1,\ldots,k-1$.

It is possible to obtain number of $3z/2(z+3)-2$ correlations from Eqs. \re{denk42}-\re{denk53} and
then to construct the system of linear equations (which has correlations and fundamental equalities given in \re{denk14}-\re{denk17}, as equations) which has
the number of $3z/2(z+3)+2$ equations. However, in order to  ensure the correctness of the correlation expressions given in Eqs. \re{denk42}-\re{denk53} we have to investigate
the spin-1/2 limits of them. It is a well known fact that the spin-1 system behaves like spin-1/2 system for large positive values of
the crystal field i.e. in the limit of $D\rightarrow \infty$. This means that in this limit, $s_i=0$ becomes as an inaccessible state for the spins on the lattice sites.
As a result of this, in this limit, all $q_i$ terms in expressions become $1$. This criteria impose that Eq. \re{denk46} can not be used for deriving that correlations. This fact is explained in Section \ref{App_b} in detail.

From Eqs. \re{denk14}-\re{denk17} and the definition of the related operators given in Eqs. \re{denk26} and \re{denk37},
we can obtain the fundamental equations of the central and perimeter spin magnetizations and quadrupolar moments as
\eq{denk54}{
\sanddr{s_0}=\summ{n=0}{z}{}\summ{p=0}{n}{}C_{np}^{(1)}\sanddr{s_\delta^{(p)}q_\delta^{(n-p)}}
} \eq{denk55}{
\sanddr{q_0}=\summ{n=0}{z}{}\summ{p=0}{n}{}C_{np}^{(2)}\sanddr{s_\delta^{(p)}q_\delta^{(n-p)}}
}
\eq{denk56}{\sanddr{s_1}=K_{0}^{(1)}+K_{1}^{(1)}\sanddr{s_0}+K_{2}^{(1)}\sanddr{q_0}}
\eq{denk57}{ \sanddr{q_1}=K_{0}^{(2)}+K_{1}^{(2)}\sanddr{s_0}+K_{2}^{(2)}\sanddr{q_0}}
where ($i=1,2$ and $j=0,1,2$)
\eq{denk58}{
\begin{array}{lcl}
C_{np}^{(i)}&=&\sandd{A_{np}}_r F_{i}(x)|_{x=0}\\
K_{j}^{(i)}&=&\sandd{B_j}_rF_i(x+\gamma)|_{x=0}.\\
\end{array}
} Configurational averages in Eq. \re{denk58} can be evaluated by the given probability
distribution of bonds $P_J\paran{J_{ij}}$ and the functions $F_{i}(x)$ are given by Eq. \re{denk10}, coefficients $A_{np}$ and $B_j$  are given by
Eqs. \re{denk27} and \re{denk38}, respectively.

We can obtain the correlations which has only perimeter sites, by using Eq. \re{denk40} in  Eqs. \re{denk42} and \re{denk43} with $k=2,3,\ldots,z$. From Eq. \re{denk42} with $m=0$ we have
\eq{denk59}{\sanddr{s_\delta^{(k)}q_\delta^{(0)}}=K_{0}^{(1)}\sanddr{s_\delta^{(k-1)}q_\delta^{(0)}}+K_{1}^{(1)}\sanddr{s_0s_\delta^{(k-1)}q_\delta^{(0)}}+
K_{2}^{(1)}\sanddr{q_0s_\delta^{(k-1)}q_\delta^{(0)}}}
and from Eq. \re{denk43} with $m=1,2,\ldots,k$ we obtain
\eq{denk60}{
\sanddr{s_\delta^{(k-m)}q_\delta^{(m)}}=K_{0}^{(2)}\sanddr{s_\delta^{(k-m)}q_\delta^{(m-1)}}+K_{1}^{(2)}\sanddr{s_0s_\delta^{(k-m)}q_\delta^{(m-1)}}+
K_{2}^{(2)}\sanddr{q_0s_\delta^{(k-m)}q_\delta^{(m-1)}}.
}

For correlations that include $s_0$ and perimeter sites, we can use operator expressions Eqs. \re{denk34} and \re{denk41}, in Eqs. \re{denk44} and  \re{denk47} respectively with
$k=1,2,\ldots,z$. From Eq. \re{denk44} we get
$$
\sanddr{s_0s_\delta^{(k)}q_\delta^{(0)}}=\paran{L_k^{(1,1)}+L_k^{(1,2)}}\sanddr{s_\delta^{(k)}q_\delta^{(0)}}+L_k^{(1,3)}\sanddr{s_\delta^{(k-1)}q_\delta^{(1)}}+
\summ{n=k}{z}{\paran{C_{n0}^{(1)}+L_{nk}^{(1,4)}}\sanddr{s_\delta^{(k)}q_\delta^{(n-k)}}}
$$
\eq{denk61}{
+\summ{n=k}{z}{L_{nk}^{(1,5)} \sanddr{s_\delta^{(k-1)}q_\delta^{(n-k+1)}}}
+\summ{n=k,}{z}{}\summ{p=2k-1}{n}{}C_{np}^{(1)}\sanddr{s_\delta^{(p-k)}q_\delta^{(n-p+k)}}
}
where $k=1,2,\ldots,z$ and from Eq. \re{denk47} with $m=1,2,\ldots,k$
\eq{denk62}{\sanddr{s_0s_\delta^{(k-m)}q_\delta^{(m)}}=(K_{0}^{(2)}+K_{2}^{(2)})\sanddr{s_0s_\delta^{(k-m)}q_\delta^{(m-1)}}+
K_{1}^{(2)}\sanddr{q_0s_\delta^{(k-m)}q_\delta^{(m-1)}}
}

Similarly, for the correlations that include $q_0$ and perimeter sites, we can use the operator expressions given in Eqs. \re{denk34} and \re{denk36}, in Eqs. \re{denk50} and \re{denk51}, respectively with $k=1,2,\ldots,z$. From Eq. \re{denk50}
$$
\sanddr{q_0s_\delta^{(k)}q_\delta^{(0)}}=\paran{L_k^{(2,1)}+L_k^{(2,2)}}\sanddr{s_\delta^{(k)}q_\delta^{(0)}}+
L_k^{(2,3)}\sanddr{s_\delta^{(k-1)}q_\delta^{(1)}}+\summ{n=k}{z}{\paran{C_{n0}^{(2)}+L_{nk}^{(2,4)}}\sanddr{s_\delta^{(k)}q_\delta^{(n-k)}}}
$$
\eq{denk63}{
+\summ{n=k}{z}{L_{nk}^{(2,5)}\sanddr{s_\delta^{(k-1)}q_\delta^{(n-k+1)}}}
+\summ{n=k,}{z}{}\summ{p=2k-1}{n}{}C_{np}^{(2)}\sanddr{s_\delta^{(p-k)}q_\delta^{(n-p+k)}}
}
and from Eq. \re{denk51} with $m=1,2,\ldots,k$
$$
\sanddr{q_0s_\delta^{(k-m)}q_\delta^{(m)}}=\paran{L_k^{(2,1)}+L_k^{(2,2)}}\sanddr{s_\delta^{(k-m)}q_\delta^{(m)}}+
L_k^{(2,3)}\sanddr{s_\delta^{(k-m+1)}q_\delta^{(m-1)}}
$$
$$
+\summ{n=k}{z}{\paran{C_{n0}^{(2)}+L_{nk}^{(2,4)}}\sanddr{s_\delta^{(k-m)}q_\delta^{(n-k+m)}}}
$$
\eq{denk64}{
+\summ{n=k}{z}{\paran{C_{n,2k-1}^{(2)}+L_{nk}^{(2,5)}} \sanddr{s_\delta^{(k-m+1)}q_\delta^{(n-k+m-1)}}}
+\summ{n=k,}{z}{}\summ{p=2k}{n}{}C_{np}^{(2)} \sanddr{s_\delta^{(p-k-m)}q_\delta^{(n-p+k+m)}}
}
where the coefficients $L_k^{(ij)}$ and $L_{nk}^{(ij)}$ are given by ($i=1,2$)

\eq{denk65}{
\begin{array}{lcl}
L_k^{(ij)}&=&\sandd{D_{k}^{(j)}}_rF_{i}(x)|_{x=0}, \quad j=1,2,3\\
L_{nk}^{(ij)}&=&\sandd{D_{nk}^{(j)}}_rF_{i}(x)|_{x=0},\quad j=4,5\\
\end{array}
}
and this completes the derivation of correlations process.

Fundamental equalities  in Eqs. \re{denk54}-\re{denk57} and Eqs. \re{denk59}-\re{denk64}
constitute linear equation system which includes number of $3/2
z(z+3)+2$ equations of correlations for the system with coordination
number $z$. By appropriately indexing the correlations, this system of equations
can be written in a matrix form and can be solved
numerically. If the system is defined by coordination number $z$
with bond, crystal field and longitudinal magnetic field probability
distributions ($P_J\paran{J_{ij}}$, $P_D\paran{D_i}$ and
$P_H\paran{H_i}$ respectively), then the functions in coefficients  can be calculated from
Eqs. \re{denk9}, \re{denk10}, the coefficients in the operators from Eqs. \re{denk27}, \re{denk35}, \re{denk38} and
finally the coefficients  in the correlation equalities from Eqs. \re{denk58}, \re{denk65} with using this
distribution functions.

Instead of deriving correlations by using Eqs. \re{denk42}, \re{denk43}, \re{denk44}, \re{denk47}, \re{denk50} and \re{denk51}; using
Eqs. \re{denk42}, \re{denk43}, \re{denk47}, \re{denk48}, \re{denk52} and \re{denk53} for correlations and
Eqs. \re{denk54}-\re{denk57} for fundamental equalities will generate another set of linear equations which
gives results that are identical to those obtained in
\cite{ref24}, which is just the extended BPA of the spin-1/2 system to the spin-1 system.

For completeness of our work, we give the DA formulation corresponding to Eqs. \re{denk54} and \re{denk55}  with the help of
Eq. \re{denk11} as
\eq{denk66}{
\sanddr{s_0}=\summ{n=0}{z}{}\summ{p=0}{n}{}C_{np}^{(1)}\sanddr{s_0}^p\sanddr{q_0}^{n-p}
} \eq{denk67}{
\sanddr{q_0}=\summ{n=0}{z}{}\summ{p=0}{n}{}C_{np}^{(2)}\sanddr{s_0}^p\sanddr{q_0}^{n-p}
} By solving the system of nonlinear equations given by  Eqs. \re{denk66}, \re{denk67}  with
the coefficients in \re{denk58} we get DA results.

\section{Results and Discussion}\label{results}

Since the aim of our work is to develop a general EFT formulation which is
capable of calculating multi site spin correlations, we give only
few results in comparison with DA and EFT approximations in the
literature. Unless otherwise stated, we mean by DA
results that the results obtained by the formulation given in Ref. \cite{ref23} and the EFT results that obtained by the formulation given in Ref. \cite{ref24}.
 The results of bond dilution problem on a honeycomb
lattice ($z=3$) and crystal field dilution problem can be found in our earlier works \cite{ref78,ref79}, respectively.

\subsection{Pure System With Zero Magnetic Field}\label{results1}

In this simplest spin-1 system, all probability distribution functions can be given by delta functions as
$P_J\paran{J_{ij}}=\delta\paran{J_{ij}-J}$, $P_D\paran{D_i}=\delta\paran{D_{i}-D}$ and
$P_H\paran{H_i}=\delta\paran{H_i}$. The functions of this system can be calculated from Eqs. \re{denk9} and \re{denk10} by using these probability distribution functions for crystal field and magnetic field and they are given as

\eq{denk68}{
F_1(x)=\frac{2\sinh{\paran{\beta x}}}{2\cosh{\paran{\beta x}}+\exp{\paran{-\beta D}}}, \quad
F_2(x)=\frac{2\cosh{\paran{\beta x}}}{2\cosh{\paran{\beta x}}+\exp{\paran{-\beta D}}}
}

Since $J_{ij}=J$ for all bonds then \re{denk58} can be written as

\eq{denk69}{C_{np}^{(k)}=\paran{\begin{array}{c}z\\n\end{array}}\paran{\begin{array}{c}n\\p\end{array}}
\left[\sinh{\paran{J\nabla}}\right]^p\left[\cosh{\paran{J\nabla}}-1\right]^{n-p} F_{k}(x)|_{x=0}}
where $k=1,2$.
We can write the hyperbolic operators in terms of the exponential operators in Eq. \re{denk69} and
by making binomial expansion of them we can obtain: \eq{denk70}{
C_{np}^{(k)}=\paran{\begin{array}{c}z\\n\end{array}}\paran{\begin{array}{c}n\\p\end{array}}
\summ{r=0}{n-p}{}\summ{i=0}{r}{}\summ{j=0}{p}{}\paran{\begin{array}{c}n-p\\r\end{array}}
\paran{\begin{array}{c}r\\i\end{array}}\paran{\begin{array}{c}p\\j\end{array}}\frac{1}{2^{r+p}}(-1)^{n-p-r+j}
\exp{[(r+p-2i-2j)J\nabla]}F_{k}(x)|_{x=0} } if we
apply the differential operator to the function $F_{k}(x)$
according to Eq. \re{denk8} we can get \eq{denk71}{
C_{np}^{(k)}=\paran{\begin{array}{c}z\\n\end{array}}\paran{\begin{array}{c}n\\p\end{array}}
\summ{r=0}{n-p}{}\summ{i=0}{r}{}\summ{j=0}{p}{}\paran{\begin{array}{c}n-p\\r\end{array}}
\paran{\begin{array}{c}r\\i\end{array}}\paran{\begin{array}{c}p\\j\end{array}}\frac{1}{2^{r+p}}(-1)^{n-p-r+j}
F_k[(r+p-2i-2j)J] }

Other coefficients for this case can be obtained by the same way from Eq. \re{denk58} as
\eq{denk72}{
\begin{array}{lcl}
K_{0}^{(k)}&=&F_k(\gamma)\\
K_{1}^{(k)}&=&1/2\left[ F_k(J+\gamma)-F_k(-J+\gamma)\right]\\
K_{2}^{(k)}&=&1/2\left[ F_k(J+\gamma)+F_k(-J+\gamma)\right]- F_k(\gamma)
\end{array}
} where $k=1,2$.

In order to construct the system of linear equations, all necessary coefficients can be obtained with the help of
Eqs. \re{denk68}, \re{denk71} and \re{denk72}. We can investigate all phase diagrams by solving the system by following the procedure given in Section \ref{model}.

\begin{figure}[h]\center
\epsfig{file=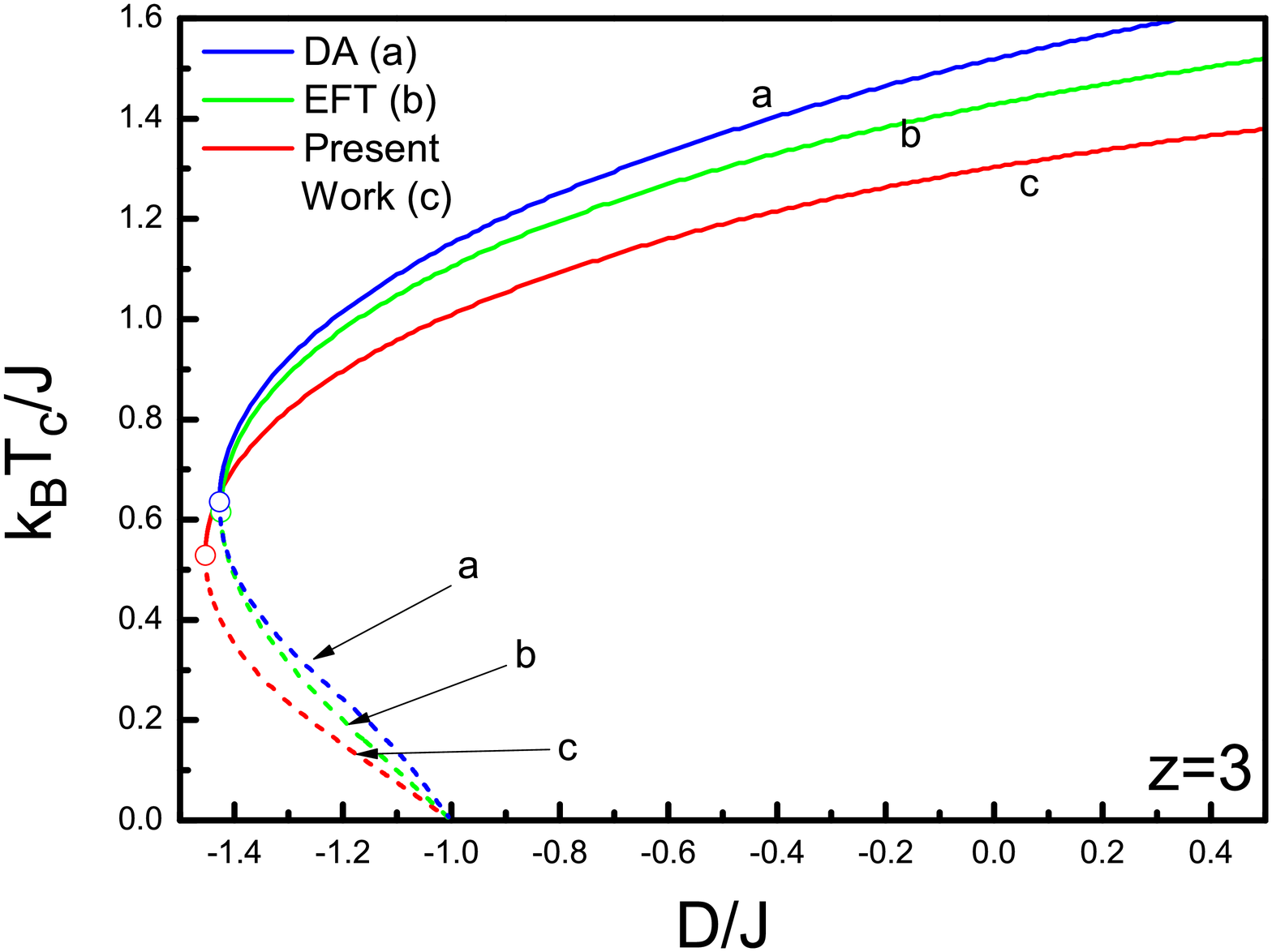, angle=0,width=8cm} \label{sek1}
%\subfigure[]{\epsfig{file=3D_T_C_D.eps, angle=0,width=8cm}}
\caption{The phase diagram of the pure system for $z=3$ in $(D/J-k_BT_c/J)$ plane. Solid
and dotted curves correspond to the second and first order phase transitions, respectively and solid circles denote the tricritical points.}
\end{figure}

The phase diagrams in $(D/J-k_BT_c/J)$ plane for
honeycomb ($z=3$) lattice can be seen in Fig. \ref{sek1} in comparison with DA and EFT. As seen in Fig. \ref{sek1} the three approximation gives
qualitatively same  phase diagrams but we should call attention to quantitative differences. While EFT gives slightly lower critical temperatures than DA, the present formulation gives lower critical temperatures than EFT for all crystal field values. The quantitative differences can be seen in
Table \ref{tablo1} for critical temperatures at $D=0$ and Table  \ref{tablo2} for tricritical points ($D_t/J,k_BT_t/J$), at which the second order phase transition line and
first order transition line meets.  The $D_t/J$ values are almost equal
for three formulations, however $k_BT_t/J$ values of DA and EFT are close to each other, while the present formulation gives lower values
for the $k_BT_t/J$. This fact can be seen in Table \ref{tablo2}. The superiority of the present formulation can also be
seen in the numerical values of critical temperatures in the limit $D\rightarrow \infty$, which are just the critical temperatures of the corresponding spin-1/2 system\cite{ref80}.

The phase diagrams in the $(D/J-k_BT_c/J)$ plane for the various three dimensional lattices ( $z=6$ simple cubic and $z=8$ body centered lattices) can be seen in Fig. \ref{sek2} in comparison with DA, as a limiting case of the bond diluted system, i.e. $c=1$. Since the quantitative and qualitative relations of the phase diagrams  of the EFT formulation and present formulation for these lattices are the same as $z=3$ phase diagram,  EFT diagrams have not been shown for this lattices.

\begin{table}[h]\label{tablo1}
%\begin{flushleft}
\begin{center}
\begin{threeparttable}
\caption{The critical temperatures of the pure system with zero magnetic field obtained by DA, EFT and the present work for the crystal field $D/J=0$.}
\renewcommand{\arraystretch}{1.3}
\begin{tabular}{llll}
\thickhline
Lattice & DA &EFT  &  Present Work \\
\hline
$3$ & 1.519& 1.428 & 1.302 \\
$4$ & 2.187& 2.114 & 1.952 \\
$6$ & 3.519& 3.466 & 3.265 \\
$8$ & 4.849& 4.809 & 4.587 \\
$12$& 7.512& 7.484 & 7.241 \\
\thickhline \\
\end{tabular}
%\end{flushleft}
\end{threeparttable}
\end{center}
\end{table}

\begin{table}[h]\label{tablo2}
%\begin{flushleft}
\begin{center}
\begin{threeparttable}
\caption{The tricritical points ($D_t/J,k_BT_t/J$) of the pure system with zero magnetic field obtained by DA, EFT and present work.}
%% The DA
%% results are obtained by the formulation given in \cite{ref23} and the EFT results are obtained by the formulation given in \cite{ref25}.
\renewcommand{\arraystretch}{1.3}
\begin{tabular}{lllllll}
Lattice & DA  & &  EFT & &Present  & Work \\
\thickhline
& $D_t$ & $k_BT_t/J$  &  $D_t$ & $k_BT_t/J$ &$D_t$ & $k_BT_t/J$\\
\hline
$3$ & 1.427& 0.632& 1.424 & 0.614 & 1.452& 0.528\\
$4$ & 1.890& 0.949 & 1.887& 0.946 & 1.912& 0.846\\
$6$ & 2.816& 1.581 & 2.814& 1.583 & 2.841& 1.454\\
$8$ & 3.741& 2.220 & 3.741& 2.217 & 3.768& 2.056\\
$12$& 5.594& 3.480 & 5.593& 3.482 & 5.621& 3.315\\
\thickhline \\
\end{tabular}
%\end{flushleft}
\end{threeparttable}
\end{center}
\end{table}

\subsection{Bond Diluted System}\label{results2}

Let us treat the bond dilution problem for homogenous crystal
fields (i.e $P_D\paran{D_i}=\delta\paran{D_i-D}$ for all $i$) and zero magnetic field (i.e $H_i=0$
for all lattice sites). For this system we assume that
the nearest-neighbor interactions are randomly distributed on the lattice
sites according to probability distribution function \eq{denk73}{
P_J\paran{J_{ij}}=\paran{1-c}\delta\paran{J_{ij}}+c\delta\paran{J_{ij}-J}
} where $c$ is the concentration of closed bonds and $0<c\le 1$.

Since the crystal field and the magnetic field distributions are same as the pure system with zero magnetic field,
the functions which are used for calculating coefficients for this system will be \re{denk68}. The configurational average in Eq. \re{denk58},
\eq{denk74}{C_{np}^{(k)}=\paran{\begin{array}{c}z\\n\end{array}}\paran{\begin{array}{c}n\\p\end{array}}
\prodd{\delta=1}{p}{}\integ{}{}{dJ_{0\delta}P_J\paran{J_{0\delta}}}\sinh{\paran{J_{0\delta}\nabla}}
\prodd{\delta=p+1}{n}{}\integ{}{}{dJ_{0\delta}P_J\paran{J_{0\delta}}}\left[\cosh{\paran{J_{0\delta}\nabla}}-1\right]
 F_{k}(x)|_{x=0}}
can be taken by using Eq. \re{denk73} in Eq. \re{denk74} and it gives
\eq{denk75}{C_{np}^{(k)}=\paran{\begin{array}{c}z\\n\end{array}}\paran{\begin{array}{c}n\\p\end{array}}
\left[c\sinh{\paran{J\nabla}}\right]^p\left[c\cosh{\paran{J\nabla}}-c\right]^{n-p}
 F_{k}(x)|_{x=0}} where $k=1,2$.

Applying the same procedure between Eqs. \re{denk69}-\re{denk71} to Eq. \re{denk75} gives
\eq{denk76}{
C_{np}^{(k)}=\paran{\begin{array}{c}z\\n\end{array}}\paran{\begin{array}{c}n\\p\end{array}}c^n
\summ{r=0}{n-p}{}\summ{i=0}{r}{}\summ{j=0}{p}{}\paran{\begin{array}{c}n-p\\r\end{array}}
\paran{\begin{array}{c}r\\i\end{array}}\paran{\begin{array}{c}p\\j\end{array}}\frac{1}{2^{r+p}}(-1)^{n-p-r+j}
F_k[(r+p-2i-2j)J].}

By the same way, the coefficients in Eq. \re{denk58} can be written for this system as
\eq{denk77}{
\begin{array}{lcl}
K_{0}^{(k)}&=&F_k(\gamma)\\
K_{1}^{(k)}&=&c/2\left[ F_k(J+\gamma)-F_k(-J+\gamma)\right]\\
K_{2}^{(k)}&=&c/2\left[ F_k(J+\gamma)+F_k(-J+\gamma)\right]- cF_k(\gamma)
\end{array}
} where $k=1,2$.

As an example of the results obtained within the present formulation in comparison with the results of DA, let us investigate the phase diagram for
simple cubic and body centered cubic lattices in $(D/J-k_BT_c/J)$ plane for the bond diluted system. The
general effect of dilution of the bonds in spin-1 BC model is to decrease the critical temperatures. At a certain bond concentration value, namely bond
percolation threshold ($c^{*}$),  ordered state is completely destroyed and the system cannot exhibit a ferromagnetic phase. Bond percolation threshold value depends on
lattice type for a given Hamiltonian and for $c<c^{*}$  system shows no phase transition at all.

\begin{figure}[h]
\subfigure[]{\epsfig{file=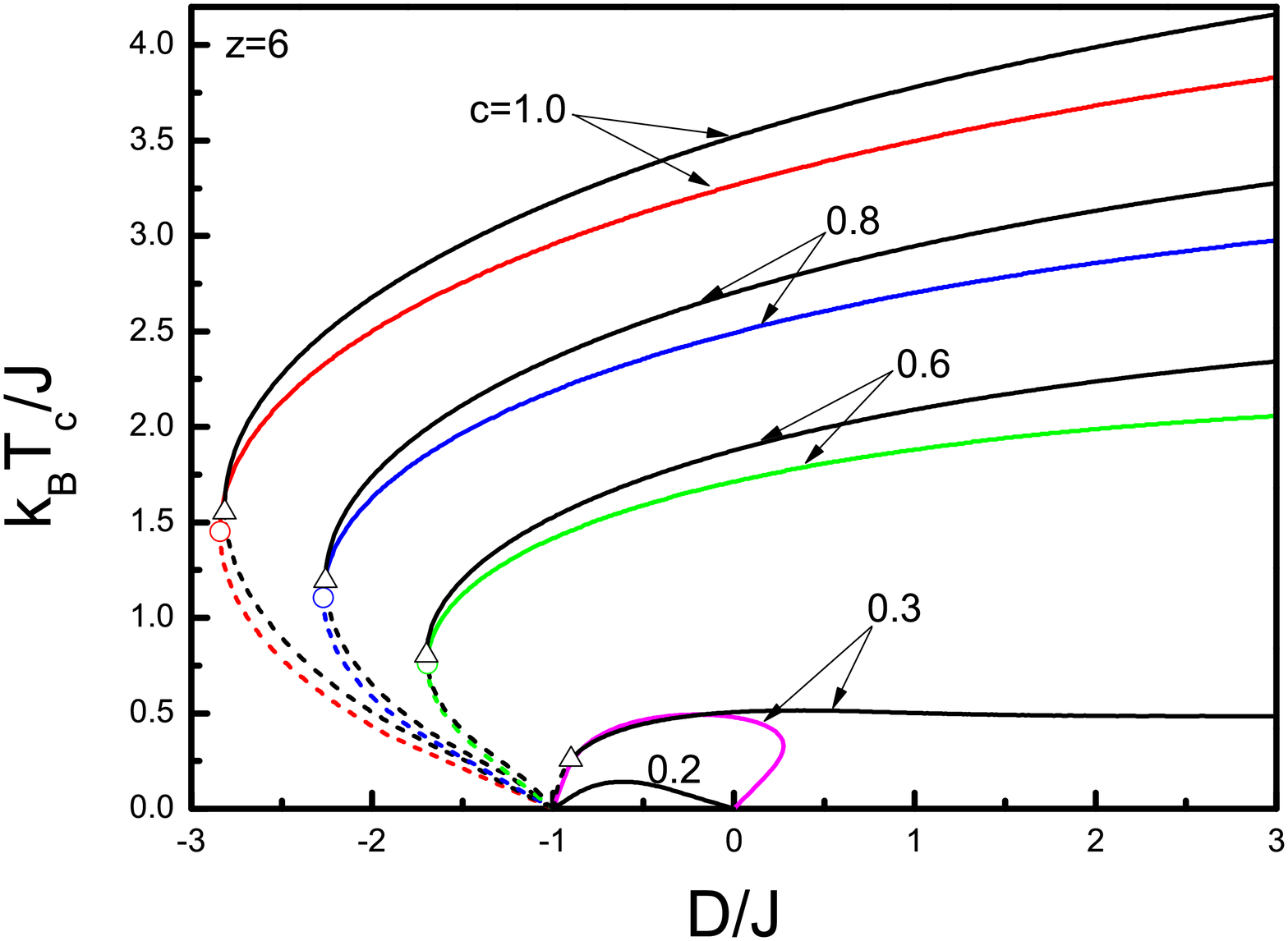, angle=0,width=8cm}}
\subfigure[]{\epsfig{file=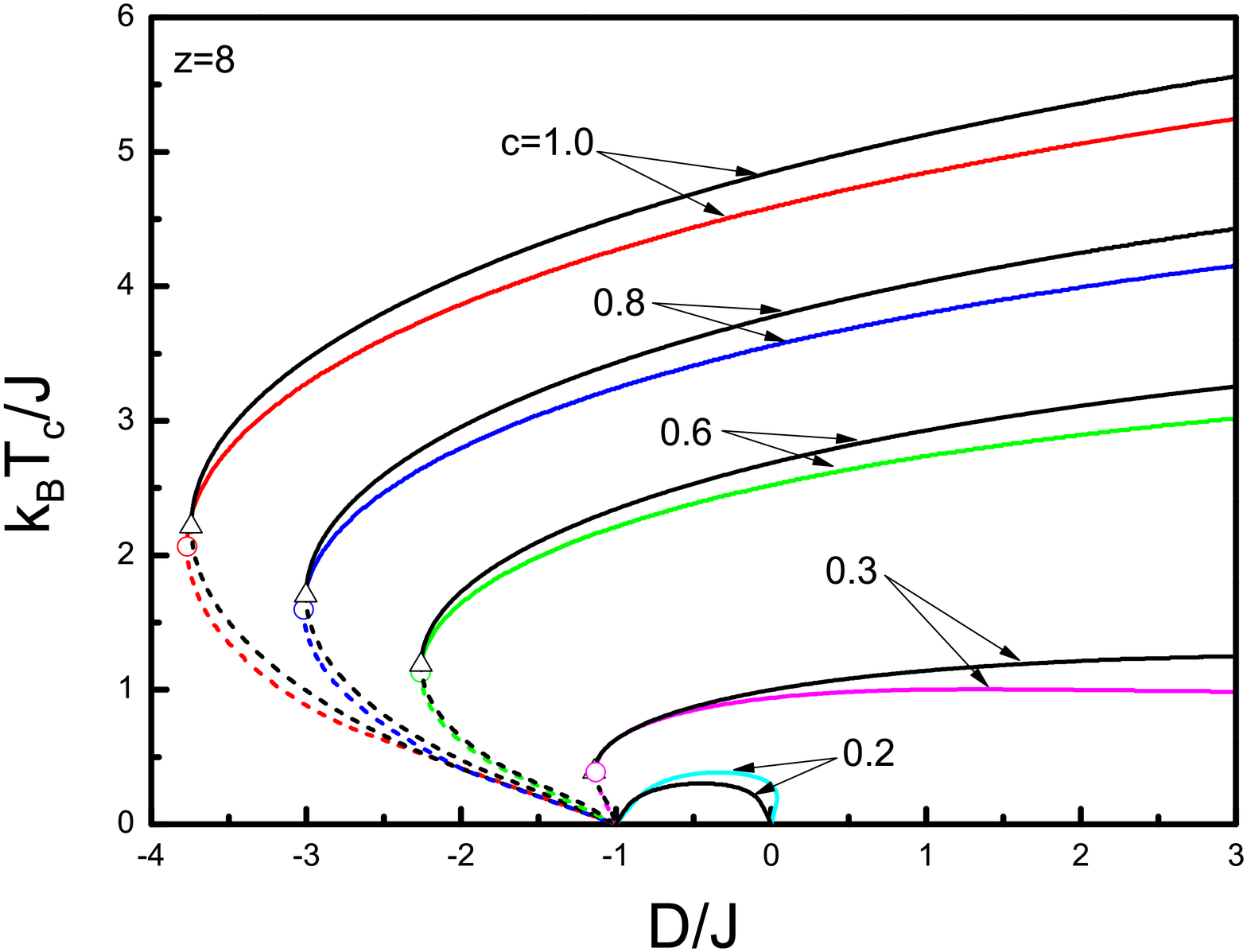, angle=0,width=8cm}}
\caption{The phase diagrams of the bond diluted system in the $(D/J-k_BT_c/J)$ plane with selected values of the bond concentration $c$. Solid
and dotted curves correspond to the second and first order phase transitions, respectively, while solid circles denote the tricritical points. For each $c$ value, black
curves are obtained by the DA  and colored curves are obtained by the
present formulation. }\label{sek2}
\end{figure}

It can be seen in Fig. \ref{sek2} that, $T_c$ values of our formulation is lower than those obtained by DA for all values of $c$, as in the pure system with zero magnetic field. This means  higher bond percolation threshold $c^{*}$ value for the present formulation with respect
to the DA. This can be seen in Table \ref{tablo3}. Since the $T_c$ values of the present formulation is closer to exact ones in comparison with DA,
it is expected that this is also true for the values of $c^{*}$. In Table \ref{tablo3} we see that, the difference in the bond
percolation threshold  values obtained in the present work and DA and EFT gets smaller as the coordination number increases.

For $D\rightarrow \infty $ it is obvious that the system behaves like spin-1/2 system. This implies that, the phase diagrams in Fig. \ref{sek2} are getting closed on the right side, after the bond percolation threshold value of the corresponding  spin-1/2 system. In general, since the calculation of the correlations within the EFT gives closer $T_c$ or $c^{*}$ value
in comparison with the exact results than DA or results of other  EFT approximations, we can say that $c$ value at
which phase diagrams getting closed on the right side in Fig. \ref{sek2} (i.e corresponding S - 1/2 bond percolation threshold values) will be closer to exact ones also. This values can be found in our earlier work \cite{ref80}.

\begin{figure}[h]\center
\epsfig{file=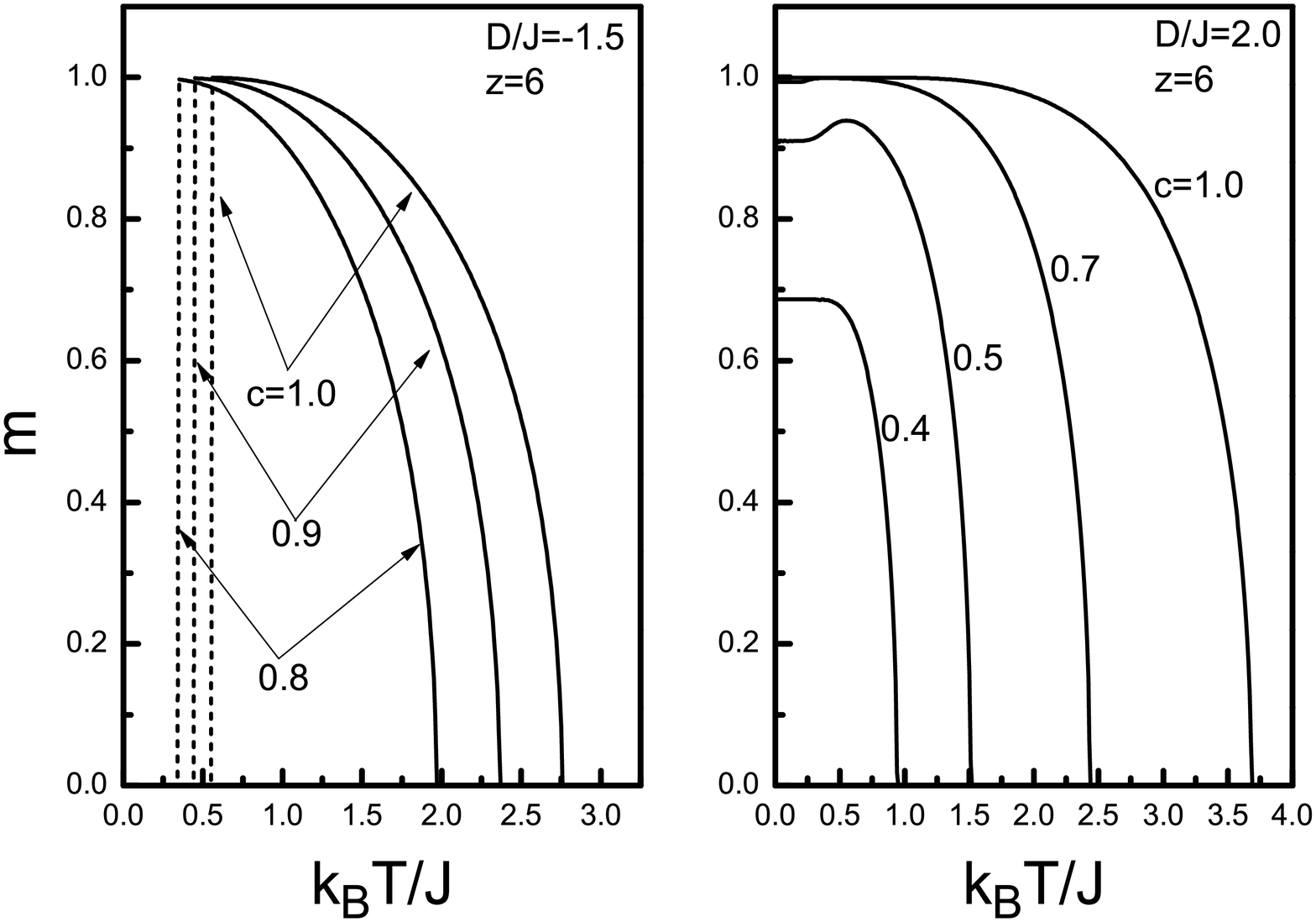, angle=0,height=6cm, width=12cm}
\caption{The variation of the magnetization with the temperature for the bond diluted simple cubic lattice, with selected values of the bond concentration $c$. }
\label{sek3}
\end{figure}

The order of the phase transition can be determined by examining the variation of the order parameter with temperature. As seen in Fig \ref{sek3}, we can conclude that all reentrant behaviors which emerge for the negative $D/J$ values are of first order.
%% in contrast to $z=3$ system \cite{ref78} (\textbf{bu kisma bir daha bak!!}).
In addition, magnetization behaviors for $D/J=2.0$ for some selected $c$ values can be seen in Fig. \ref{sek3}.

\begin{figure}[h]\center
\epsfig{file=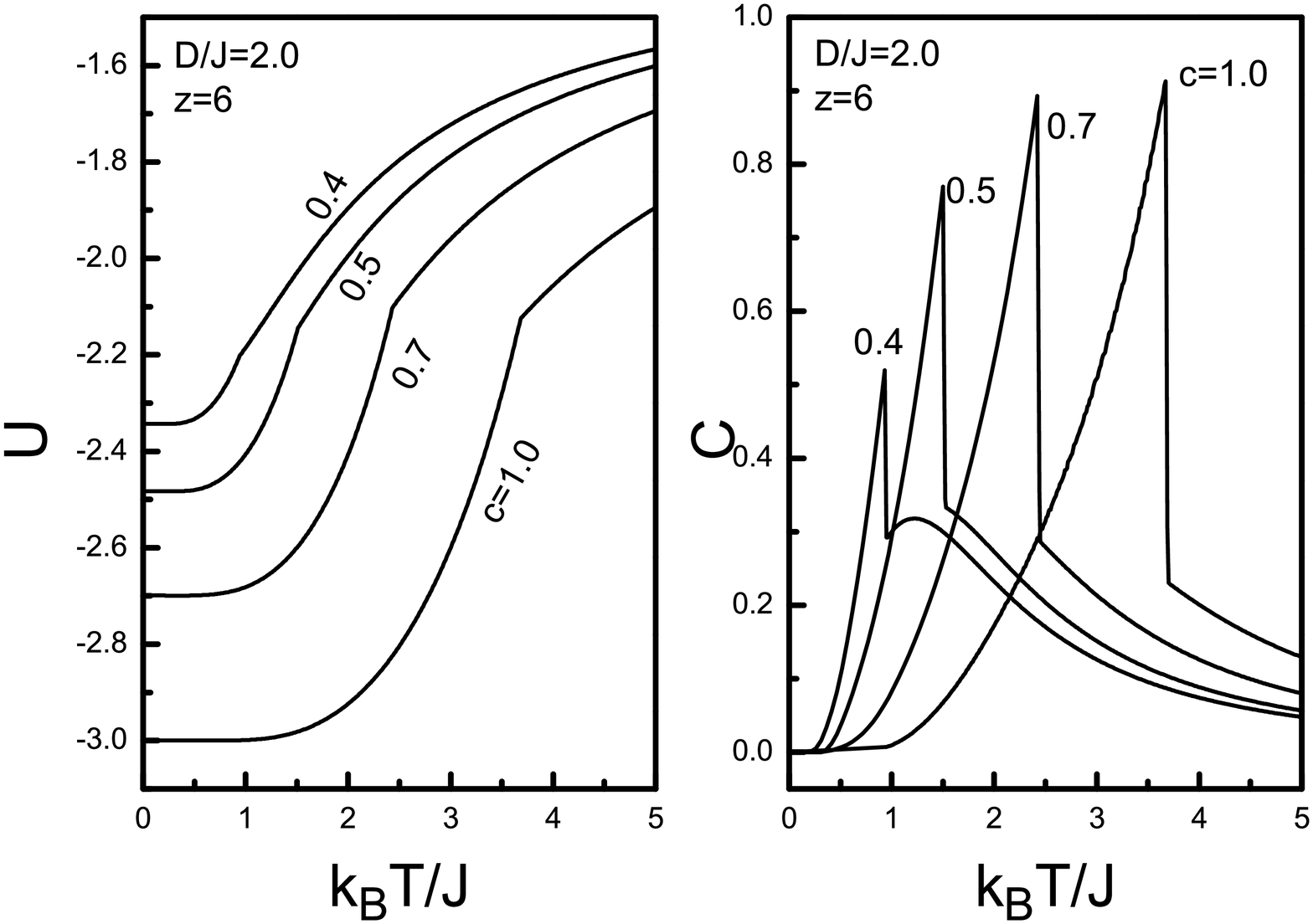, angle=0,height=6cm,width=12cm}
\caption{The variation of the internal energy ($U$) and the specific heat ($C$) with the temperature for the bond diluted simple cubic lattice, with selected values of the bond concentration $c$. }\label{sek4}
\end{figure}

The other superiority of the present formulation is computability of the some thermodynamic functions which are
related to the correlations (e.g. internal energy and thus specific heat).  The internal energy and specific heat
of simple cubic lattice for $D/J=2.0$ with some selected bond concentration values are plotted in Fig. \ref{sek4}. The behaviors of
the internal energy and specific heat are more reasonable than DA results, since DA neglects all multi site spin correlations. Besides it gives zero internal energy after the critical temperature, which is physically impossible. One important point about the specific heat curves is, appearance of a hump as $c$ decreases. While $c$ decreases, peak of the
specific heat gets smoother and gives place to a growing hump after the bond
percolation threshold value $c^*$.

\begin{table}[h]\label{tablo3}
%\begin{flushleft}
\begin{center}
\begin{threeparttable}
\caption{The bond percolation threshold values ($c^{*}$)  of the bond diluted system with zero magnetic field, obtained by DA,EFT and present formulation for different lattices.}
\renewcommand{\arraystretch}{1.3}
\begin{tabular}{llll}
\thickhline
Lattice & DA  &EFT &  Present Work \\
\hline
$3$ & 0.333 & 0.347 & 0.391 \\
$4$ & 0.249 & 0.256 & 0.282 \\
$6$ &  0.166 & 0.168 & 0.181 \\
$8$ & 0.124 & 0.126 & 0.133 \\
$12$& 0.083 & 0.084 & 0.086 \\
\thickhline \\
\end{tabular}
%\end{flushleft}
\end{threeparttable}
\end{center}
\end{table}

%$3$ & 0.34-0.33 & 0.35-0.34 & 0.40-0.39\\
%$4$ & 0.26 - 0.25 & 0.26-0.25 & 0.29 -0.28 \\
%$6$ & 0.17 - 0.16 & 0.17 - 0.16 & 0.19- 0.18 \\
%$8$ & 0.13 - 0.12 & 0.13 - 0.12 & 0.14 - 0.13 \\
%$12$& 0.09 - 0.08 & 0.09 - 0.08 & 0.09 - 0.08 \\

%$3$ & 0.333 & 0.347 & 0.391 \\
%$4$ & 0.249 & 0.256 & 0.282 \\
%$6$ &  0.166 & 0.168 & 0.181 \\
%$8$ & 0.124 & 0.126 & 0.133 \\
%$12$& 0.083 & 0.084 & 0.086 \\

%\begin{figure}[h]
%\epsfig{file=bd_perc_tum.ps, angle=0,width=8cm}\label{sek4}
%\epsfig{file=bd_perc_sag.ps, angle=0,width=8cm}
%\caption{}
%\end{figure}

\subsection{Crystal Field Dilution Problem}\label{results3}
As a final investigation let us investigate the system with diluted crystal fields. The random crystal fields are distributed to lattice sites according to given
probability distribution function,
\eq{denk78}{
P_D\paran{D_{i}}=\paran{1-p}\delta\paran{D_{i}}+p\delta\paran{D_{i}-D}
}The system is under zero magnetic field and there is no bond dilution, i.e  $H_i=0$ and  $J_{ij}=J$ for all $i,j$. This changes only the functions defined in Eq. \re{denk10}. Using $P_H\paran{H_i}=\delta\paran{H_i}$ and Eq. \re{denk78} in Eq. \re{denk10} will give

\eq{denk79}{
F_1(x)=\frac{2(1-p)\sinh{\paran{\beta x}}}{2\cosh{\paran{\beta x}}+1}+\frac{2p\sinh{\paran{\beta x}}}{2\cosh{\paran{\beta x}}+\exp{\paran{-\beta D}}}}

\eq{denk80}{
F_2(x)=\frac{2(1-p)\cosh{\paran{\beta x}}}{2\cosh{\paran{\beta x}}+1}+\frac{2p\cosh{\paran{\beta x}}}{2\cosh{\paran{\beta x}}+\exp{\paran{-\beta D}}}
}

By using Eqs. \re{denk76} and \re{denk77} with $c=1$ and Eqs. \re{denk79} and \re{denk80} we can construct the system of linear equations.

The distribution function given in Eq. \re{denk78} distributes crystal fields in $p$ percentage of lattice sites $D$ and remaining $1-p$ percentage of lattice sites zero. Because of this, $p=0$ phase diagrams
in  $(D/J-k_BT_c/J)$ plane will be the line parallel to the $D/J$ axis and the value of  the $k_BT_c/J$  is just the critical temperature of the pure system
(i.e. system with no external magnetic field, homogenous distributed $D$ and $J$ ) at $D=0$.
The phase diagrams in this plane will evolve towards the phase diagram of pure system in the same plane,
when concentration value goes to $p=1$. During this evolution, up to a certain $p=p^\star$ value, phase
diagrams will not intersect $D/J$ axis on the left side of the $(D/J-k_BT_c/J)$ plane
 (i.e. region with negative $D/J$ values). This means that, system is in the ordered phase at low temperatures for all $D/J$ values for $p<p^\star$.

\begin{figure}[h]
\subfigure[]{\epsfig{file=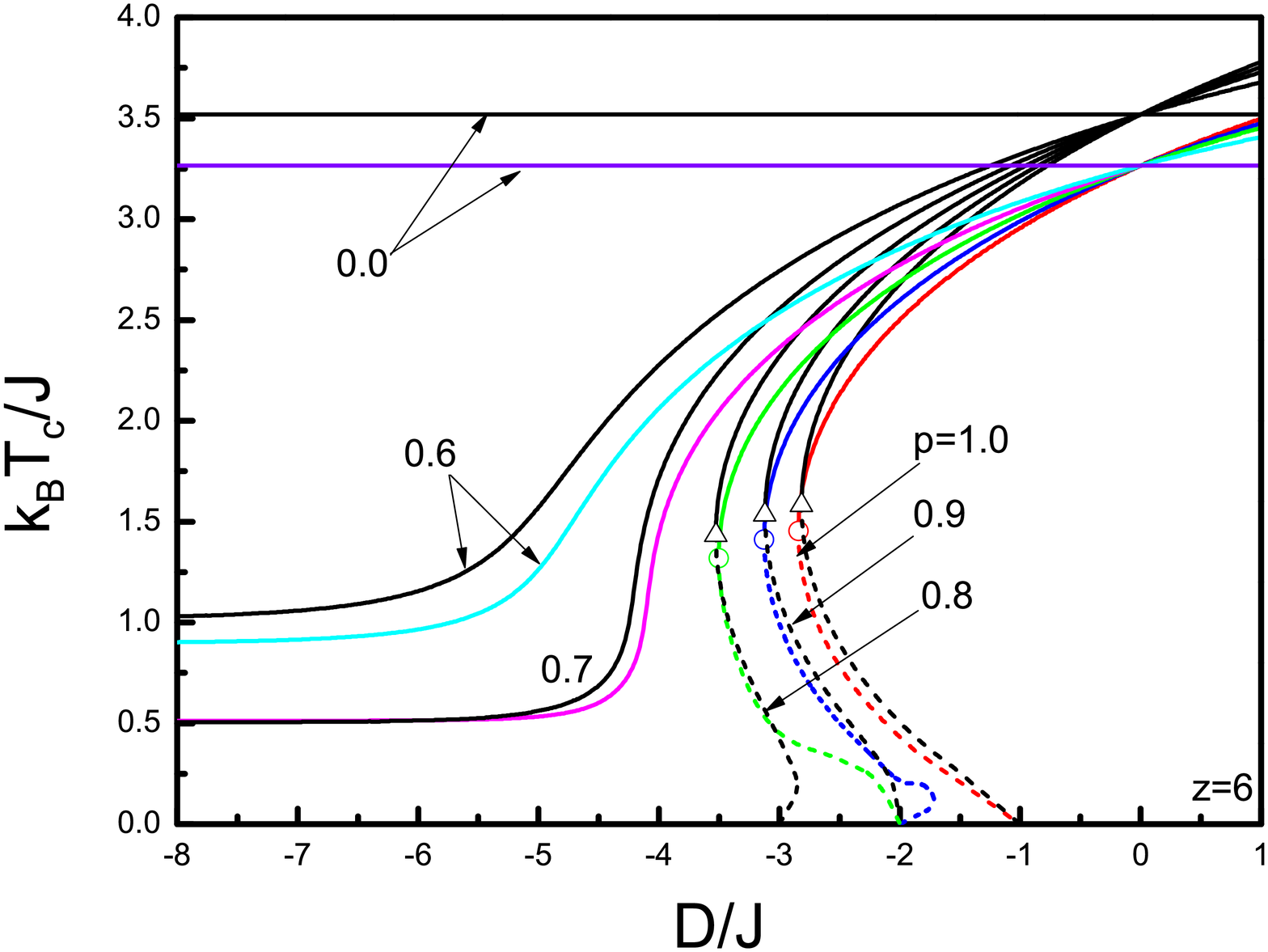, angle=0,width=8cm}}
\subfigure[]{\epsfig{file=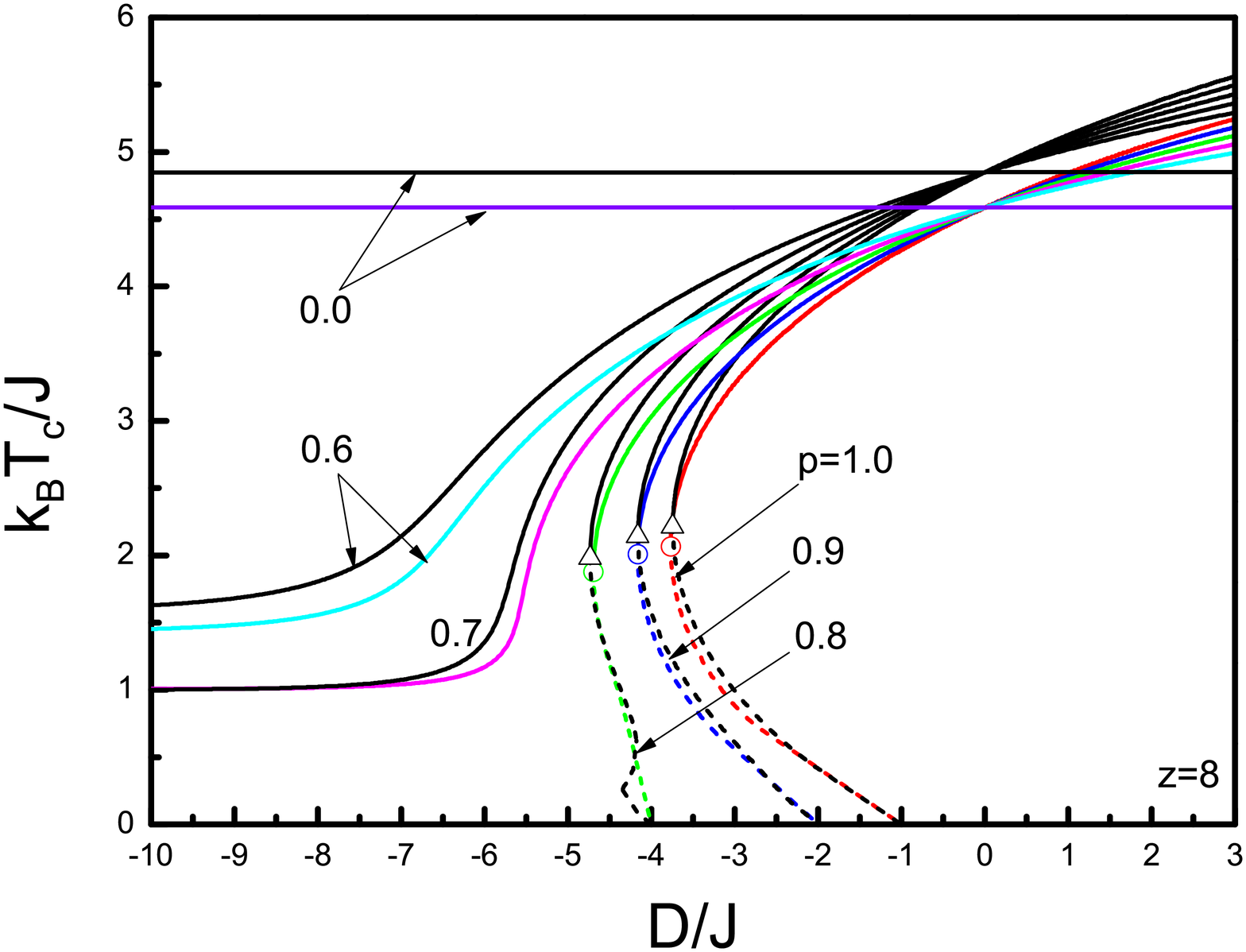, angle=0,width=8cm} }
\caption{The phase diagrams of the crystal field diluted system in $(D/J-k_BT_c/J)$ plane with selected values of the crystal field dilution parameter ($p$). Solid
and dotted curves correspond to the second and first order phase transitions, respectively and solid circles denote the tricritical points. For each $p$ value, black curves are obtained by the DA  and colored curves are obtained by the
present formulation. }\label{sek5}
\end{figure}

The phase diagrams of the lattices with coordination numbers $z=6,8$ (simple cubic and body centered lattice respectively) can be seen in Fig. \ref{sek5} in comparison
with DA phase diagrams. Again we can see from Fig. \ref{sek5} that the $T_c$ values of introduced formulation are lower than those obtained by DA. This makes the critical concentration value $p^\star$ lower than that of DA. The $p^\star$ values for different lattices can be seen in Table \ref{tablo4}.

\begin{figure}[h]\center
\epsfig{file=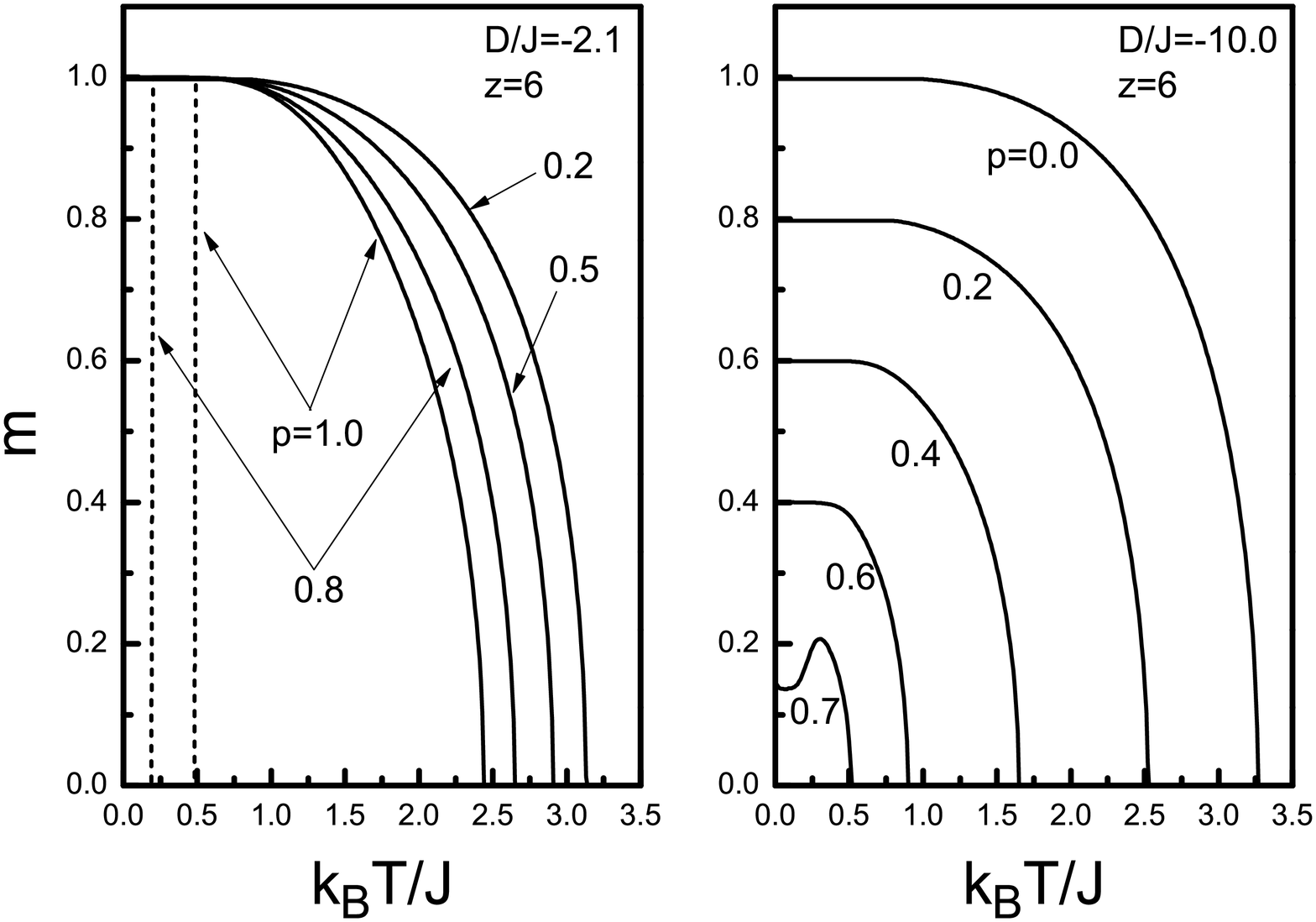, angle=0,height=6cm, width=12cm}
\caption{The variation of the magnetization with temperature for the system on a simple cubic lattice,  with selected values of the crystal field dilution parameter $p$.}
\label{sek6}
\end{figure}

\begin{table}[h]\label{tablo4}
%\begin{flushleft}
\begin{center}
\begin{threeparttable}
\caption{The critical crystal field concentration values  of crystal field diluted system with zero magnetic field, obtained by DA,EFT and present formulation for different lattices.}
\renewcommand{\arraystretch}{1.3}
\begin{tabular}{llll}
\thickhline
Lattice & DA  &EFT &  Present Work \\
\hline
$3$ & 0.484& 0.476  & 0.379 \\
$4$ & 0.604& 0.611 & 0.544 \\
$6$ & 0.729& 0.745 & 0.718 \\
$8$ & 0.794& 0.810 & 0.802 \\
$12$& 0.861& 0.875 & 0.892 \\
\thickhline \\
\end{tabular}
%\end{flushleft}
\end{threeparttable}
\end{center}
\end{table}

%\begin{figure}[h]
%\epsfig{file=crys_dil_c.ps, angle=0,width=8cm}\label{sek1}
%\caption{}
%\end{figure}

\section{Conclusion}\label{conclusion}

In this work we have presented a general formulation for the spin-1 Ising system with BC model with nearest neighbor interactions. Within the introduced formulation
we can calculate the multi site spin correlations in a representative manner. This capability shows its superiority in the results of the critical temperatures and
other critical values (e.g. bond
percolation threshold values) and behavior of thermodynamic functions (e.g specific heat) as a function of the temperature. Our formulation can be applied
to the systems with quenched disordered systems. These type of disorders in real materials can be extremely important since they can induce important macroscopic effects on the thermal and magnetic properties of the material. The systems corresponding to these systems cannot be solved exactly in most cases. Hence, some approximations must be done. But, including approximations in the calculations means that
loosing accuracy of the results. This is a disadvantageous situation from the experimental point of view of the  material science, although making approximations simplify the solutions. Thus, it is desired
to improve approximations as much as possible to give more closer solutions to exact ones.

As shown in Section \ref{results}, our formulation gives the best approximated solutions of BC model for the spin-1 system within the framework of EFT. Besides, it can give more reasonable
behaviors of some thermodynamic functions (e.g. internal energy, specific heat) in comparison to the other EFT formulations.

On the other hand we believe that, it is very important to develop a general formulation which is capable of applying  to the related systems with arbitrary coordination numbers. In conclusion, we hope that the formulation and results obtained in this work may be beneficial
form both theoretical and experimental point of view.

\section{Appendix A}\label{App_a}

During the derivation process of correlations in a sequential order
given in Eq. \re{denk21} we envision that each coefficient $A_{np}$ in the
operators moves to another correlation as a multiplier of it at
each step of the derivation. Thus, derivation process generate a movement
of the coefficients $A_{np}$. We can track the 'path' all coefficients
during this derivation process and determine the place of all coefficients at
the  $k_{th}$ step of this movement.

According to  Eq. \re{denk28} there are three possible 'ways' for a
coefficient. Let us call movements in these three possible different
ways as movements in regime 1,regime 2 and regime 3. In the regime 1,
movement goes on according to the first line of the right hand side
of Eq. \re{denk28}. Similarly in the regime 2 movement goes as the  the
second line of the right hand side of Eq. \re{denk28} and the movement
in regime 3 goes as third line.

In Eq. \re{denk29}, we give the path of the coefficient $A_{00}$  from
the beginning to the end of the movement. This movement always stays
in the regime 1 since at each step, the step number $k$ satisfies
$k>l+m$, where indices $l,m$ defines the term as
$s_\delta^{(l)}q_\delta^{(m)}$ which is the starting point of the $k_{th}$ step.

Now we give the paths of the remaining coefficients according to this reasoning. At the
beginning of the movement, the coefficients $A_{n0}$ ($n\ne 0$) are
placed at the $s_\delta^{(0)}q_\delta^{(n)}$ . For the first step
($k=1$) since $0<k\le n$ the movement starts in the regime 2. From
Eq. \re{denk28} we can generate the path of these coefficients while the
derivation process in a sequential order given in Eq. \re{denk21}. The first $n$
step of these coefficients in the regime 2 are given as follows \eq{denka1}{
s_\delta^{(0)}q_\delta^{(n)}\rightarrow
s_\delta^{(1)}q_\delta^{(n-1)} \rightarrow \ldots \rightarrow
s_\delta^{(n)}q_\delta^{(0)} } We see from Eq. \re{denka1} that at the end of the first $n$ step, the
coefficient $A_{n0}$ arrives at $s_\delta^{(n)}q_\delta^{(0)}$. After
then, the movement goes on in the regime 1 since the step $k=n$ starts to
fulfill the condition $k> n$ \eq{denka2}{
s_\delta^{(n)}q_\delta^{(0)}\rightarrow
s_\delta^{(n+1)}q_\delta^{(0)} \rightarrow \ldots \rightarrow \ldots
\rightarrow s_\delta^{(z)}q_\delta^{(0)} }

Thus for the coefficient $A_{n0}$, we can conclude that for the steps $k\le n$ it
is at the place $s_\delta^{(k)}q_\delta^{(n-k)}$ (Eq. \re{denka1}),
and for the steps $k> n$ it is at the place
$s_\delta^{(k)}q_\delta^{(0)}$ (Eq.  \re{denka2}) i.e we arrive
Eq. \re{denk31}.

Remaining coefficients $A_{np}$ ($n,p>0$) are placed at
$s_\delta^{(p)}q_\delta^{(n-p)}$ at the beginning.  Since for the
first step $k=1\le n$ is satisfied, then the coefficient $A_{np}$ starts its movement
in the regime 3 and it stays in this regime until the step  $r$
which is $p-r=r$ i.e. $r=p/2$ according to Eq. \re{denk28}. \eq{denka3}{
s_\delta^{(p)}q_\delta^{(n-p)}\rightarrow
s_\delta^{(p-1)}q_\delta^{(n-p+1)}  \ldots \rightarrow
s_\delta^{(p-r)}q_\delta^{(n-p+r)} } If the index $p$, which is the
number of $s_\delta$ in the term $s_\delta^{(p)}q_\delta^{(n-p)}$ is
odd, then we have $r=(p+1)/2$. Thus we can say that, the
coefficient $A_{np}$  ($n,p>0$) is at  $s_\delta^{(p-k)}q_\delta^{(n-p+k)}$,
for the end of the step $k\le p/2$ (for even $p$)  and for the end of the step $k\le (p+1)/2$ (for odd $p$)
 i.e. the first lines
of the right hand sides of Eqs. \re{denk32} and \re{denk33}.

At the end of the movement given in Eq. \re{denka3} the coefficient $A_{np}$
($n,p>0$) is at the place $s_\delta^{(p-r)}q_\delta^{(n-p+r)}$.
After than the movement goes in the second regime since the step number
$k$ becomes greater than $p-r$. The movement stays in the regime
2 for $t$ steps as,

\eq{denka4}{ s_\delta^{(p-r)}q_\delta^{(n-p+r)}\rightarrow
s_\delta^{(p-r+1)}q_\delta^{(n-p+r-1)}  \ldots \rightarrow
s_\delta^{(p-r+t)}q_\delta^{(n-p+r-t)} }

Since in the regime 3
and regime 2  keeps the places of the coefficients
$s_\delta^{(l)}q_\delta^{(m)}$ to be $l+m$ constant, at a certain
step, step number will catch $l+m$ then the movement goes on a regime
1 according to the first line of the right hand side of Eq. \re{denk28}.
This step number satisfies $r+t=n$. Then while for even $p$ this
condition reads $p/2+t=n$, for odd $p$ reads $(p+1)/2+t=n$. We find
that for even $p$, $t=n-p/2$ and for odd $p$, $t=n-(p+1)/2$. Thus
the movement given in Eq. \re{denka4} ends with $s_\delta^{(n)}q_\delta^{(0)}$ for even $p$
and $s_\delta^{(n-1)}q_\delta^{(1)}$ for odd $p$. We conclude from
Eq. \re{denka4} that for even $p$ the coefficient $A_{np}$ ($n,p>0$) is
at the place $s_\delta^{(k)}q_\delta^{(n-k)}$ and for the odd $p$ at the place
$s_\delta^{(k-1)}q_\delta^{(n-k+1)}$ for the end of the step $r<k\le r+t$. These results are the second
lines of the right hand sides of Eqs. \re{denk32} and \re{denk33}.

Now the movement is in the regime 1 and according to Eq. \re{denk28} it
goes as
\eq{denka5}{ s_\delta^{(n)}q_\delta^{(0)}\rightarrow
s_\delta^{(n+1)}q_\delta^{(0)}  \ldots \rightarrow
s_\delta^{(z)}q_\delta^{(0)} } \eq{denka6}{
s_\delta^{(n-1)}q_\delta^{(1)}\rightarrow
s_\delta^{(n)}q_\delta^{(1)}  \ldots \rightarrow
s_\delta^{(z)}q_\delta^{(1)} } From Eqs. \re{denka5} and \re{denka6} we
can say that the coefficient $A_{np}$ ($n,p>0$) is at the place
$s_\delta^{(k)}q_\delta^{(0)}$ for even $p$ and at the place
$s_\delta^{(k-1)}q_\delta^{(1)}$ for odd $p$ for the end of the step $k> n$. This corresponds to the third
lines of the right hand sides of the Eqs. \re{denk32} and \re{denk33} and this
completes our derivation of Eqs. \re{denk31}-\re{denk33}.

\section{Appendix B}\label{App_b}

As explained in Section \ref{model}, Eqs. \re{denk42}-\re{denk53} have to give correct spin-1/2 equivalents in the limits $D\rightarrow \infty$ and $q_i\rightarrow 1$ where $i=0,1,\ldots,z$.

The central spin and the perimeter spin averages for the nearest neighbor
spin-1/2 Ising model with random bond and random magnetic field
distribution can be given as\cite{ref80}

\eq{denkc1}{\sanddr{s_0}=\sanddr{\Theta_0^{(1/2)}}f(x)|_{x=0}}
\eq{denkc2}{\sanddr{s_1}=\sanddr{\Phi_0^{(1/2)}}f(x+\gamma)|_{x=0} }
where
\eq{denkc3}{
g(x,H_i)=\tanh{\left[\beta\paran{x+H_i}\right]}, \quad f(x)=\integ{}{}{dH_i P_H\paran{H_i}g\paran{x,H_i}
}}
The operators in Eqs. \re{denkc1} and \re{denkc2} can be written as
\eq{denkc4}{\Theta_0^{(1/2)}=\summ{n=0}{z}{A^{(1/2)}_ns_\delta^{(n)}}}
\eq{denkc5}{\Phi_0^{(1/2)}=B_0^{(1/2)}+B_1^{(1/2)}s_0} where the coefficients are given by
\eq{denkc6}{
\begin{array}{lcl}
A^{(1/2)}_n&=&\paran{\begin{array}{c}z\\n\end{array}}\prodd{\delta=1}{n}{\sinh{\paran{J_{0\delta}\nabla}}}
\prodd{\delta=n+1}{z}{\cosh{\paran{J_{0\delta}\nabla}}},\quad n=0,1,\ldots z \\
B_0^{(1/2)}&=&\cosh{\paran{J_{0\delta}\nabla}}\\
B_1^{(1/2)}&=&\sinh{\paran{J_{0\delta}\nabla}}
\end{array}
}

With the operators
\eq{denkc7}{\Phi_k^{(1/2)}=B_0^{(1/2)}s_\delta^{(k)}+B_1^{(1/2)}s_0s_\delta^{(k)}}
\eq{denkc8}{s_0\Phi_k^{(1/2)}=B_0^{(1/2)}s_0s_\delta^{(k)}+B_1^{(1/2)}s_\delta^{(k)}}
\eq{denkc9}{\Theta_k^{(1/2)}=s_\delta^{(k-1)}\summ{n=0}{k-1}{A_{2n+1}^{(1/2)} }+s_\delta^{(k)}\summ{n=0}{k}{A_{2n}^{(1/2)} }+\summ{n=2k+1}{z}{A_{n}^{(1/2)} s_\delta^{(n-k)}}}
we can obtain spin-1/2 multi spin correlations as
\eq{denkc10}{\sanddr{s_\delta^{(k)}}=\sanddr{\Phi_{k-1}^{(1/2)}}f(x+\gamma)|_{x=0}}
\eq{denkc11}{\sanddr{s_0s_\delta^{(k)}}=\sanddr{\Theta_k^{(1/2)}}f(x)|_{x=0}}
\eq{denkc12}{\sanddr{s_0s_\delta^{(k)}}=\sanddr{s_0\Phi_{k-1}^{(1/2)}}f(x+\gamma)|_{x=0}}

Now, we can see from Eq. \re{denk9}
\eq{denkc13}{\lim_{D\rightarrow \infty}G_1(x,H_i)=g(x,H_i), \quad \lim_{D\rightarrow \infty}G_2(x,H_i)=1}
and then
\eq{denkc14}{\lim_{D\rightarrow \infty}F_1(x)=f(x), \quad \lim_{D\rightarrow \infty}F_2(x)=1}
where $f(x),g(x)$ are the spin-1/2 functions which are defined in Eq. \re{denkc3}.

Let us label all operators and coefficients related to the spin-1 system with superscript $(1)$. With a little combinatorics one can show that
\eq{denkc15}{\summ{n=p}{z}{A_{np}^{(1)}}=A_p^{(1/2)}} is valid for
the relation between the coefficients of spin-1/2 and spin-1 Ising system which is given in Eq. \re{denk27}.

Let us look at the $q_i\rightarrow 1,\quad (i=0,1,\ldots, z)$ limits of all necessary operators for deriving multi site correlations and fundamental equalities of spin-1 Ising system which are given in Eqs. \re{denk26}, \re{denk34}, \re{denk36}, \re{denk40} and \re{denk41}. We start with Eq. \re{denk26}.

\eq{denkc16}{ \lim_{q_i\rightarrow
1}\Theta_{00}^{(1)}=\summ{n=0}{z}{}\summ{p=0}{n}{}A_{np}^{(1)}s_\delta^{(p)}=
\summ{p=0}{z}{s_\delta^{(p)}}\summ{n=p}{z}{}A_{np}^{(1)} } According
to Eq. \re{denkc15} this will give  Eq. \re{denkc4}. Thus \eq{denkc17}{
\lim_{q_i\rightarrow 1}\Theta_{00}^{(1)}=\Theta_{0}^{(1/2)} } The
spin-1/2 limit of the operator given in Eq. \re{denk34} by using Eq. \re{denk35} can be obtained as
$$
\lim_{q_i\rightarrow
1}\Theta_{k,0}^{(1)}=\paran{\summ{n=0}{z}{A_{n0}^{(1)}}+\summ{n=1}{k-1}{}\summ{p=2}{n^{\prime\prime}}{}A_{np}^{(1)}+
\summ{n=k}{z}{\summ{p=2}{2k-1^{\prime\prime}}{}A_{np}^{(1)}}}s_\delta^{(k)}
$$
\eq{denkc18}{
+\paran{\summ{n=1}{k-1}{}\summ{p=1}{n^{\prime}}{}A_{np}^{(1)}+\summ{n=k}{z}{\summ{p=1}{2k-2^{\prime}}{}A_{np}^{(1)}}}s_\delta^{(k-1)}
+\summ{n=k,}{z}{}\summ{p=2k-1}{n}{}A_{np}^{(1)}s_\delta^{(p-k)}
}
by rearranging the sums we get \eq{denkc19}{ \lim_{q_i\rightarrow
1}\Theta_{k,0}^{(1)}=
s_\delta^{(k)}\summ{p=0}{2k^{\prime\prime}}{}\summ{n=p}{z}{}A_{np}^{(1)}+
s_\delta^{(k-1)}\summ{p=1}{2k-1^{\prime}}{}\summ{n=p}{z}{}A_{np}^{(1)}+
\summ{p=2k+1}{z}{}\summ{n=p}{z}{}A_{np}^{(1)}s_\delta^{(p-k)} }
According to Eq. \re{denkc15}, the right hand side of Eq. \re{denkc19} is equal
to the right hand side of Eq. \re{denkc9}, then \eq{denkc20}{
\lim_{q_i\rightarrow 1}\Theta_{k,0}^{(1)}=\Theta_{k}^{(1/2)} }

The spin-1/2 limit of the operator given in Eq. \re{denk36} is

$$
\lim_{q_i\rightarrow 1}\Theta_{k,m}^{(1)}=\paran{\summ{n=0}{z}{A_{n0}^{(1)}}+\summ{n=1}{k-1}{}\summ{p=2}{n^{\prime\prime}}{}A_{np}^{(1)}+
\summ{n=k}{z}{\summ{p=2}{2k-1^{\prime\prime}}{}A_{np}^{(1)}}}s_\delta^{(k-m)}
$$
\eq{denkc21}{
+\paran{\summ{n=1}{k-1}{}\summ{p=1}{n^{\prime}}{}A_{np}^{(1)}
+\summ{n=k}{z}{A_{n,2k-1}^{(1)}}
+\summ{n=k}{z}{\summ{p=1}{2k-2^{\prime}}{}A_{np}^{(1)}}
}s_\delta^{(k-m+1)}+\summ{n=k,}{z}{}\summ{p=2k}{n}{}A_{np}^{(1)} s_\delta^{(p-k-m)}
}
We can see from Eq. \re{denkc9} that, $\Theta_{k-m}$  contains $s_\delta^{(k-m-1)}$, instead of  $s_\delta^{(k-m+1)}$ in Eq. \re{denkc21}, then we can conclude that
\eq{denkc22}{
\lim_{q_i\rightarrow 1}\Theta_{k,m}^{(1)}\ne \Theta_{k-m}^{(1/2)}
}

The spin-1/2 limit of Eq. \re{denk40} is given by
\eq{denkc23}{\lim_{q_i\rightarrow 1}\Phi_{k,m}^{(1)}=\paran{B_0^{(1)}+B_{2}^{(1)}}s_\delta^{(k-m)}+B_{1}^{(1)}s_0s_\delta^{(k-m)}}
With the help of  Eq. \re{denk38}, we can conclude from  Eq. \re{denkc23} with comparing \re{denkc7} that
\eq{denkc24}{
\lim_{q_i\rightarrow 1}\Phi_{k,m}^{(1)}= \Phi_{k-m}^{(1/2)}
}
Similarly we can obtain the results for the operators given in Eq. \re{denk41} as
\eq{denkc25}{
\begin{array}{lcl}
\lim_{q_i\rightarrow 1}s_0\Phi_{k,m}^{(1)}&=& s_0\Phi_{k-m}^{(1/2)}\\
\lim_{q_i\rightarrow 1}q_0\Phi_{k,m}^{(1)}&=& \Phi_{k-m}^{(1/2)}\\
\end{array}
}

Now we can look at the  spin-1/2 limits of expressions
given in Eqs. \re{denk42}-\re{denk53}, in order to decide whether they have
correct limits or not. Let us start with Eq. \re{denk42}. Whereas the
limit of the left hand side of Eq. \re{denk42} is \eq{denkc26}{
\lim_{q_i\rightarrow 1}\sanddr{s_\delta^{(k-m)}q_\delta^{(m)}}
=\sanddr{s_\delta^{(k-m)}} } the limit of the right hand side of
Eq. \re{denk42} can obtained as by using Eqs. \re{denkc14} and \re{denkc24}
\eq{denkc27}{ \lim_{q_i\rightarrow
1}\sanddr{\Phi_{k-1,m}^{(1)}}F_1\paran{x+\gamma}|_{x=0}=
\sanddr{\Phi_{k-1-m}^{(1/2)}}f\paran{x+\gamma}|_{x=0} }  Then with the help of Eqs. \re{denkc26} and \re{denkc27}, we can
obtain for \re{denk42}
\eq{denkc28}{\sanddr{s_\delta^{(k-m)}}=\sanddr{\Phi_{k-1-m}^{(1/2)}}f\paran{x+\gamma}|_{x=0}}
This is nothing but Eq. \re{denkc10} for the correlation
$\sanddr{s_\delta^{(k-m)}}$ for spin-1/2 system. Thus we can conclude
that expression given in Eq. \re{denk42} has correct spin-1/2 limit.

In a similar manner, Eq. \re{denk43} has spin-1/2 limit as \eq{denkc29}{
\sanddr{s_\delta^{(k-m)}} = \sanddr{\Phi_{k-m}^{(1/2)}}1 } From
Eq. \re{denkc7} we can write Eq. \re{denkc29} as \eq{denkc30}{
\sanddr{s_\delta^{(k-m)}} = \sanddr{s_\delta^{(k-m)}}B_0^{(1/2)}1 +
\sanddr{s_0s_\delta^{(k-m)}}B_1^{(1/2)}1} and from Eq. \re{denkc6}, we have
\eq{denkc31}{ \sanddr{s_\delta^{(k-m)}} =
\sanddr{s_\delta^{(k-m)}}} i.e tautology.

If we continue to work on the spin-1/2 limits of the expressions given in
Eqs. \re{denk44}-\re{denk53}, we can see that the expressions
\re{denk44},\re{denk45}, \re{denk48}, \re{denk49} and \re{denk53}
gives the correct limits as in \re{denkc28} for the expression
\re{denk42} and the expressions \re{denk47}, 
\re{denk50}, \re{denk51} and \re{denk52}  gives tautology as in \re{denkc31} for
the expression \re{denk43}. The only expression gives false spin-1/2
limit is \re{denk46}.

\end{document}